\documentclass[manuscript]{aastex}
\usepackage{cases}
\usepackage{arydshln}
\usepackage{color}

\shorttitle{d'Alembert-type scheme for few-body problem}
\shortauthors{Y. Minesaki}
\begin{document}
\title{
d'Alembert-type scheme with a chain regularization for $N$-body problem
}
\author{Yukitaka Minesaki}
\affil{Tokushima Bunri University, Nishihama, Yamashiro-cho, Tokushima
770-8514, Japan}
\begin{abstract}
We design an accurate orbital integration scheme for the general $N$-body
problem preserving all the conserved quantities but the angular momentum.
This scheme is based on the chain concept \citep{MA} and is regarded as
an extension of a d'Alembert-type scheme \citep{Betsch2005}
for constrained Hamiltonian systems.
It also coincides with the discrete-time general three-body problem
\citep{Minesaki-2013a} for particle number $N = 3$.
Although the proposed scheme is only second-order accurate, it can
accurately reproduce some periodic orbits, which generic geometric
numerical integrators cannot do.
\end{abstract}
\section{Introduction}
To find periodic orbits in the $N$-body problem ($N \ge 3$), we use
methods (e.g., \cite{Baltagiannis-b, Broucke-1969}) consisting of
two procedures:
(1) We introduce a rotating-pulsating frame where a few primaries are
fixed.
(2) We use the Runge--Kutta--Fehlberg method and set the allowable energy
variation and errors of the positions, or the Steffensen method with
recurrent power series so that we compute the periodic orbits in one period.
However, these methods are not suitable for reproducing the orbits in
this problem for a long time interval because of the following two drawbacks: (a) A
rotating-pulsating frame, in which the methods described in (2) are
applied, has to be altered in accordance with periodic orbits.
(b) The methods in (2) cannot accurately compute periodic orbits for a
long time interval because they do not preserve any conserved quantities.
\par
On the other hand, for any initial condition, including the conditions of
some periodic orbits, numerical integration methods are applied to the
$N$-body problem in the barycentric inertial frame.
If we use a non-geometric integration method, this method cannot reproduce periodic orbits for a long time interval because of drawback (b).
In addition, even if we used each of the geometric integration methods
(e.g., the symplectic and energy-momentum methods), they cannot necessarily
reproduce periodic orbits.
Both the symplectic and energy-momentum methods cannot illustrate
elliptic orbits in the two-body problem
\citep{Minesaki-2002,Minesaki-2004} and elliptic Lagrange orbits in the
three-body problem \citep{Minesaki-2013a}.
To overcome drawbacks (a) and (b), the author already proposed
the discrete-time general three-body problem (d-G$3$BP)
\citep{Minesaki-2013a} and the discrete-time restricted three-body
problem (d-R$3$BP) \citep{Minesaki-2013c} for the general three-body
problem (G$3$BP) and restricted three-body problem (R$3$BP) in the
barycentric inertial frame, respectively.
These schemes \citep{Minesaki-2013a, Minesaki-2013c} are given by an
extension of a d'Alembert-type scheme \citep{Betsch2005}.
The d-G$3$BP retains all the conserved quantities but the angular momentum,
and the d-R$3$BP preserves all the conserved quantities but the Jacobi
integration.
In this paper, we design an accurate orbital integration scheme like the d-G$3$BP and d-R$3$BP for the
general $N$-body problem (G$N$BP).
The new scheme is based on a d'Alembert-type scheme
\citep{Betsch2005} and a chain regularization \citep{MA}.
It keeps all the conserved quantities except the angular momentum and can accurately
compute some periodic orbits.
\par
This paper is organized as follows.
In Section 2, after labeling the masses according to the chain concept
\citep{MA} and using the Levi-Civita transformation \citep{Levi-Civita}, we
express the general $N$-body problem as a constrained Hamiltonian system
without Lagrangian multipliers.
Further, we rewrite this problem using only the vectors related to the
chained ones.
In Section 3, we apply the same discrete-time formulation adopted for
the G$3$BP in \citep{Minesaki-2013a} to the resulting problem, so we have
a discrete-time problem.
We prove that the discrete-time problem preserves all the conserved
quantities of the G$N$BP except the angular momentum.
In Section 4, we check that the discrete-time problem ensures such
preservation of the general $N$-body problem numerically.
Moreover, we show that it correctly calculates some periodic orbits.
\section{Regularization of General $N$-body Problem}
For an arbitrary number of masses $N$, we give the transformation
formulae, equations of motion for the G$N$BP, and selection of a chain of
interparticle vectors such that the close encounters requiring
regularization are included in the chain.
This formulation includes the same transformation formulae and selection
of a chain as in \citep{MA}.
It has the advantage that its computational cost is far lower than that
of Heggie's global formulation \citep{Heggie} for a large number of masses $N$.
\par
In Section 2.1, we briefly review the G$N$BP in the barycentric frame
and how to form a chain of interparticle vectors and label masses
using the chain algorithm in \citep{MA}.
In Section 2.2, using the Levi-Civita transformation \citep{Levi-Civita},
we rewrite the G$N$BP, which is similar to the problem given by Heggie's
global regularization \citep{Heggie}.
For a large number of masses $N$, the rewritten problem involves many
redundant variables.
In Section 2.3, using some constraints, we express the problem in terms of
only the chained position and momentum vectors to reduce the number of redundant
variables.
\subsection{Labeling Particles Using Chain Concept}
The small distance between two bodies experiencing a close
encounter is represented as a difference between large numbers in
straightforward formulations of the $N$-body problem.
Thus, round-off easily becomes a significant source of error.
To avoid this, we use the chain concept of \citep{MA}
introduced for regularization algorithms.
\par
In this chain method, a chain of interparticle vectors is constructed so
that all the particles are included in this chain.
Note that small distances are part of the chain.
We begin by searching for the shortest distance, which is taken as the first
piece of the chain.
Next, we find the particle closest to one or the other end of the presently
known part of the chain.
Then, we add this particle to the end of the chain that is closer.
This process is repeated until all the particles are involved.
After every integration step, we check whether any non-chained
vector is shorter than the smallest of the chained vectors that are in
contact with one or the other end of the vector under consideration, namely,
if any triangle formed by two consecutive chain vectors has the shortest
side non-chained.
If this is the case, a new chain is formed.
Hereafter, suppose the masses are relabeled $1$, $2$, $\cdots$, $N$
along the chain.
\par
We assume that $\mathbf{q}_i' \equiv (q'_{i[1]}, q'_{i[2]})$ is the position
vector of a point with mass $m_i$ in the barycentric frame.
We also define $\mathbf{p}'_i \equiv (p'_{i[1]}, p'_{i[2]})$ as a
momentum conjugate to $\mathbf{q}'_i$.
We set the gravitational constant equal to one for simplicity.
In addition, $N$ position vectors $\mathbf{q}'_1, \cdots, \mathbf{q}'_N$
satisfy the following constraints:
\begin{eqnarray*}
\displaystyle \sum_{i=1}^N m_i \mathbf{q}'_i = \mathbf{0}.
\end{eqnarray*}
\par
The equations of motion in the barycentric frame are given by the
Hamiltonian:
\begin{eqnarray}
\displaystyle
H = \frac{1}{2}
\sum_{i=1}^N
\frac{|\mathbf{p}'_i|^2}{m_i} \!
- \sum_{i=1}^{N-1} \sum_{j=i+1}^{N}
\frac{m_i m_j}{|\mathbf{q}'_i \!-\! \mathbf{q}'_j|}.
\label{Sec2.1:CQ.H}
\end{eqnarray}
The dynamical system corresponding to this Hamiltonian is
\begin{eqnarray}
\displaystyle
\frac{d}{dt} \mathbf{q}_i' = \frac{1}{m_i} \mathbf{p}_i', \quad
\frac{d}{dt} \mathbf{p}_i'
= m_i \left(
\sum_{k=1}^{i-1} \frac{m_k (\mathbf{q}_{k}' - \mathbf{q}_{i}')}
{|\mathbf{q}_{k}' - \mathbf{q}_{i}'|^3}
- \sum_{k=i+1}^N \frac{m_k
(\mathbf{q}_{i}'-\mathbf{q}_{k}')}{|\mathbf{q}_{i}'- \mathbf{q}_{k}'|^3}
\right), \ 1 \le i \le N.
\label{Sec2.1:CQ.diff2-q}
\end{eqnarray}
However, for two-body close encounters, we need to
simultaneously use two position vectors in the barycentric frame.
Therefore, the barycentric frame is not useful for computing close
encounters between two masses.
\subsection{General $N$-body Problem with Redundant Variables}
The G$N$BP in the relative frame is much more
symmetric than that in the barycentric frame.
It also has a significant advantage in investigating such properties as
periodic orbits and close encounters
(e.g., \citep{Broucke-Relative,Broucke-Periodic,Explicit-Symplectic}).
It can be integrated numerically without catastrophic errors after the
Levi-Civita or Kustaanheimo--Stiefel transformation
\citep{KS,Levi-Civita,SS}.
\par
In Section 2.2.1, we rewrite the G$N$BP in the relative frame.
The resulting problem involves very many gravitational force terms
for a large number of masses $N$.
Thus, we deform this system to reduce the number of force terms.
In Section 2.2.2, we rewrite the system using the Levi-Civita variables.
\subsubsection{General $N$-body Problem in Relative Frame}
We introduce a relative frame to consider two-body close approaches easily.
We use the relative position vectors
$\mathbf{q}_{ij} \equiv (q_{ij[1]}, q_{ij[2]})$ defined by
\begin{eqnarray}
\displaystyle
\mathbf{q}_{ij} = \mathbf{q}'_i - \mathbf{q}'_j,\quad
1 \le i < j \le N,
\label{Sec2.2:RC.q_{i,j}}
\end{eqnarray}
and the momentum $\mathbf{p}_{ij} \equiv (p_{ij[1]},p_{ij[2]})$
conjugate to $\mathbf{q}_{ij}$ as
\begin{eqnarray}
\displaystyle
\mathbf{p}_{ij} = \frac{m_j \mathbf{p}'_i - m_i \mathbf{p}'_j}{m},
\quad 1 \le i < j \le N, \label{Sec2.2:RC.p_{i,j}}
\end{eqnarray}
where $m$ is the total mass, $\sum_{i=1}^N m_i$, of the G$N$BP.
These position and momentum vectors also satisfy the following constraints:
\begin{subnumcases}
{
}
\displaystyle
\mbox{\boldmath $\phi$}_{1jk} (\mathbf{q})
\equiv \left( \phi_{1jk[1]} (\mathbf{q}), \phi_{1jk[2]} (\mathbf{q}) \right)
\equiv
\mathbf{q}_{1j} + \mathbf{q}_{jk} - \mathbf{q}_{1k} = \mathbf{0},
\quad 2 \le j < k \le N, & \label{Sec2.2:RC.Cst-q} \\
\displaystyle
\mbox{\boldmath $\psi$}_{ijk} (\mathbf{p})
\equiv \left( \psi_{ijk[1]} (\mathbf{p}), \psi_{ijk[2]} (\mathbf{p}) \right)
\equiv \frac{\mathbf{p}_{ij}}{m_i m_j}
+ \frac{\mathbf{p}_{jk}}{m_j m_k}
- \frac{\mathbf{p}_{ik}}{m_i m_k} = \mathbf{0}, \quad 1 \le i < j < k \le N.
& \label{Sec2.2:RC.Cst-p}
\end{subnumcases}
Here,
{\footnotesize
$\mathbf{q} =
(\!\!
\begin{array}{c:c:c|c:c:c|c|c}
q_{1,2[1]}, q_{1,2[2]}, & \cdots & q_{1,N[1]}, q_{1,N[2]}, &
q_{2,3[1]}, q_{2,3[2]}, & \cdots & q_{2,N[1]}, q_{2,N[2]}, &
\cdots & q_{N-1,N[1]}, q_{N-1,N[2]}
\end{array}
\!\!)
$}
$\in \mathbb{R}^{N(N-1)}$
and
\ \\
{\footnotesize
$\mathbf{p} =
(\!\!
\begin{array}{c:c:c|c:c:c|c|c}
p_{1,2[1]}, p_{1,2[2]}, & \cdots & p_{1,N[1]}, p_{1,N[2]}, &
p_{2,3[1]}, p_{2,3[2]}, & \cdots & p_{2,N[1]}, p_{2,N[2]}, &
\cdots & p_{N-1,N[1]}, p_{N-1,N[2]}
\end{array}
\!\!)
$}
$\in \mathbb{R}^{N(N-1)}$.
\par
To obtain the inverse transformations of equations (\ref{Sec2.2:RC.q_{i,j}}) and
(\ref{Sec2.2:RC.p_{i,j}}), we have to solve system
(\ref{Sec2.2:RC.q_{i,j}}) for the position vectors $\mathbf{q}'_i$ and
system (\ref{Sec2.2:RC.p_{i,j}}) for the momentum vectors $\mathbf{p}'_i$.
Unfortunately, no vector is uniquely determined.
However, if we choose
\begin{eqnarray}
&& \displaystyle
\mathbf{q}'_i = \frac{1}{m}
\left( \sum_{j=i+1}^N m_j \mathbf{q}_{ij} - \sum_{j=1}^{i-1} m_j
 \mathbf{q}_{ji} \right), \
\mathbf{p}'_i = \sum_{j=i+1}^N \mathbf{p}_{ij}
- \sum_{j=1}^{i-1} \mathbf{p}_{ji},\
\quad 1 \le i \le N
\label{Sec2.2:qp_i},
\end{eqnarray}
then these relations follow equations (\ref{Sec2.2:RC.q_{i,j}}) and
(\ref{Sec2.2:RC.p_{i,j}}).
Accordingly, we adopt equation (\ref{Sec2.2:qp_i}) as a transformation from the
relative frame to the barycentric frame.
\par
Substitution of transformations (\ref{Sec2.2:RC.q_{i,j}}) and
(\ref{Sec2.2:RC.p_{i,j}}) and their time differentiations into
equation (\ref{Sec2.1:CQ.diff2-q}) yields the following system:
\begin{subnumcases}
{
}
\displaystyle
\frac{d}{dt} \mathbf{q}_{ij} = \frac{m}{m_i m_j} \mathbf{p}_{ij}, \quad
1 \le i < j \le N, & \label{Sec3.1:Lagrange-type.Cn-Eq-q} \\
\displaystyle
\frac{d}{dt} \mathbf{p}_{1j} = - m_1 m_j
\frac{\mathbf{q}_{1j}}{|\mathbf{q}_{1j}|^3}
+ \sum_{k=2}^{j-1} \mbox{\boldmath $f$}_{kj} (\mathbf{q})
- \sum_{k=j+1}^N \mbox{\boldmath $f$}_{jk} (\mathbf{q}), \quad 2 \le j
 \le N,
& \label{Sec3.1:Lagrange-type.Cn-Eq-p1} \\
\displaystyle
\frac{d}{dt} \mathbf{p}_{ij} = - m_i m_j
\frac{\mathbf{q}_{ij}}{|\mathbf{q}_{ij}|^3}
- \mbox{\boldmath $f$}_{ij} (\mathbf{q}), \quad
2 \le i < j \le N,
& \label{Sec3.1:Lagrange-type.Cn-Eq-p}
\end{subnumcases}
where $\mbox{\boldmath $f$}_{ij} (\mathbf{q})$ is a function
of a vector $\mathbf{q}$ defined by
\begin{eqnarray}
\displaystyle \mbox{\boldmath $f$}_{ij} (\mathbf{q})
\!\!&\equiv&\!\! - \sum_{k=1}^{i-1} \frac{m_i m_j m_k}{m}
\left(
\frac{\mathbf{q}_{ki}}{|\mathbf{q}_{ki}|^3}
+ \frac{\mathbf{q}_{ij}}{|\mathbf{q}_{ij}|^3}
- \frac{\mathbf{q}_{kj}}{|\mathbf{q}_{kj}|^3}
\right)
+ \sum_{k=i+1}^{j-1} \frac{m_i m_j m_k}{m}
\left(
\frac{\mathbf{q}_{ik}}{|\mathbf{q}_{ik}|^3}
+ \frac{\mathbf{q}_{kj}}{|\mathbf{q}_{kj}|^3}
- \frac{\mathbf{q}_{ij}}{|\mathbf{q}_{ij}|^3} \right)
\nonumber\\
&& \displaystyle - \sum_{k=j+1}^N \frac{m_i m_j m_k}{m}
\left(
\frac{\mathbf{q}_{ij}}{|\mathbf{q}_{ij}|^3}
+ \frac{\mathbf{q}_{jk}}{|\mathbf{q}_{jk}|^3}
- \frac{\mathbf{q}_{ik}}{|\mathbf{q}_{ik}|^3}
\right), \quad 2 \le i < j \le N.
\label{def-GTBP-lambda}
\end{eqnarray}
We can regard the system composed of equations (\ref{Sec2.2:RC.Cst-q}) and (7)
as the G$N$BP in the relative frame.
\par
Through equation (\ref{Sec2.2:qp_i}), the Hamiltonian (\ref{Sec2.1:CQ.H}) leads
to
\begin{eqnarray*}
\displaystyle
H = \sum_{i=1}^{N-1} \sum_{j=i+1}^N
\left( \frac{m}{2 m_i m_j} |\mathbf{p}_{ij}|^2
- \frac{m_i m_j}{|\mathbf{q}_{ij}|} \right)
+ \frac{1}{2} \sum_{i=1}^{N-2} \sum_{j=i+1}^{N-1} \sum_{k=j+1}^N
m_i m_j m_k
| \mbox{\boldmath $\psi$}_{ijk} (\mathbf{p}) |^2.
\end{eqnarray*}
However, it does not yield system (7), so it cannot be regarded as
a Hamiltonian.
System (7) is governed by the following Hamiltonian:
\begin{eqnarray}
H_{\mbox{\footnotesize rel}}
\equiv \sum_{i=1}^{N-1} \sum_{j=i+1}^N
\left( \frac{m}{2 m_i m_j} |\mathbf{p}_{ij}|^2
- \frac{m_i m_j}{|\mathbf{q}_{ij}|} \right)
+ \sum_{j=2}^{N-1} \sum_{k=j+1}^N \mbox{\boldmath $\phi$}_{1jk}
(\mathbf{q}) \cdot \mbox{\boldmath $\lambda$}_{jk},
\label{H_rel}
\end{eqnarray}
where $\mbox{\boldmath $\lambda$}_{jk} = (\lambda_{jk[1]}, \lambda_{jk[2]})$
are Lagrange multipliers
\footnote{
Using the second time derivative of the constraints
(\ref{Sec2.2:RC.Cst-q}), namely,
$d \left( \mathbf{p}_{1j}/m_1 m_j + \mathbf{p}_{jk}/m_j m_k - \mathbf{p}_{1k}/m_1
m_k \right)/dt = \mathbf{0}$ $(2 \le j < k \le N)$, we can give the Lagrange
multipliers as $\mbox{\boldmath $\lambda$}_{jk} = \mathbf{f}_{jk}
(\mathbf{q})$.
Therefore, system (7) is governed by (\ref{H_rel}).}.
The values of $H$ and $H_{\mbox{\footnotesize rel}}$ coincide with
\begin{eqnarray}
h_{\mbox{\footnotesize rel}} \equiv
 \sum_{i=1}^{N-1} \sum_{j=i+1}^N
\left( \frac{m}{2 m_i m_j} |\mathbf{p}_{ij}|^2
- \frac{m_i m_j}{|\mathbf{q}_{ij}|} \right).
\label{h_rel}
\end{eqnarray}
\subsubsection{General $N$-body Problem with Levi-Civita Variables}
In the numerical integration of the G$N$BP, multibody close encounters
result in serious numerical difficulties due to the errors associated with
singularities in the G$N$BP.
The Levi-Civita regularization is a standard technique for removing the
singularities.
It combines a time regularization with the Levi-Civita transformation
\citep{Levi-Civita}.
In this subsection, using the Levi-Civita variables \citep{Levi-Civita},
we rewrite the G$N$BP (7).
\par
We apply the Levi-Civita transformation \citep{Levi-Civita} to the
vectors $\mathbf{q}_{ij}$, obtaining the vectors
$\mathbf{Q}_{ij} \equiv \left( Q_{ij[1]},Q_{ij[2]} \right)$.
The relations between $\mathbf{q}_{ij}$ and $\mathbf{Q}_{ij}$ are given
by
\begin{eqnarray}
\displaystyle
\mathbf{q}_{ij} = \mathbf{Q}_{ij} \mathbf{L} (\mathbf{Q}_{ij})^{\top},
\quad 1 \le i < j \le N,
\label{Sec4.1:Levi-Civita}
\end{eqnarray}
where the Levi-Civita matrix \citep{Levi-Civita} is defined as
\begin{eqnarray*}
\displaystyle
\mathbf{L} (\mathbf{Q}_{ij}) \equiv \left[
\begin{array}{cc}
Q_{ij[1]} & - Q_{ij[2]} \\
Q_{ij[2]} & Q_{ij[1]}
\end{array}
\right].
\end{eqnarray*}
New momentum vectors $\mathbf{P}_{ij} \equiv \left( P_{ij[1]},P_{ij[2]}
\right)$ are also related to the old ones $\mathbf{p}_{ij}$
by the equations
\begin{eqnarray}
\displaystyle
\mathbf{p}_{ij}
= \frac{1}{2 |\mathbf{Q}_{ij}|^2} \mathbf{P}_{ij}
\mathbf{L} (\mathbf{Q}_{ij})^{\top} ,
\ 1 \le i < j \le N.
\label{Sec4.1:Levi-Civita-p}
\end{eqnarray}
Theoretically, we can use the solution of relations
(\ref{Sec4.1:Levi-Civita}) and (\ref{Sec4.1:Levi-Civita-p}) to obtain
$\mathbf{Q}_{ij}$ and $\mathbf{P}_{ij}$ from $\mathbf{q}_{ij}$ and
$\mathbf{p}_{ij}$.
In numerical computation, to avoid cancellation of significant digits,
we compute $\mathbf{Q}_{ij}$ and $\mathbf{P}_{ij}$ for $1 \le i < j \le
N$ as follows:
\begin{eqnarray}
&& \hspace*{-10mm}
\mathbf{Q}_{ij}
\equiv \left\{
\begin{array}{cc}
\displaystyle \left(
\frac{q_{ij[2]}}{\sqrt{- 2 q_{ij[1]} + 2 |\mathbf{q}_{ij}|}},
\frac{1}{2} \sqrt{- 2 q_{ij[1]} + 2 |\mathbf{q}_{ij}|}
\right) & \mbox{if} \ q_{ij[1]} < 0, \\
\displaystyle \left(
\frac{1}{2} \sqrt{2 q_{ij[1]} + 2 |\mathbf{q}_{ij}|},
\frac{q_{ij[2]}}{\sqrt{2 q_{ij[1]} + 2 |\mathbf{q}_{ij}|}}
\right) & \mbox{if} \ q_{ij[1]} \ge 0,
\end{array}
\right. \label{q->Q} \\
&& \hspace*{-10mm}
\mathbf{P}_{ij}
\equiv \left\{\!\!\!\!
\begin{array}{cc}
\displaystyle \left(\!
\frac{2 \!\left(\!
p_{ij[1]} q_{ij[2]} \!-\! p_{ij[2]} q_{ij[1]} \!+\! p_{ij[2]} |\mathbf{q}_{ij}|
\!\right)}{\sqrt{- 2 q_{ij[1]} + 2 |\mathbf{q}_{ij}|}}, \
\frac{2 \left(\!
p_{ij[1]} q_{ij[1]} \!+\! p_{ij[2]} q_{ij[2]} \!-\! p_{ij[1]} |\mathbf{q}_{ij}|
\!\right)}{\sqrt{- 2 q_{ij[1]} + 2 |\mathbf{q}_{ij}|}}
\!\right) & \mbox{if} \ q_{ij[1]} < 0, \\
\displaystyle \left(\!
\frac{2 \left(\!
p_{ij[1]} q_{ij[1]} \!+\! p_{ij[2]} q_{ij[2]} \!+\! p_{ij[1]} |\mathbf{q}_{ij}|
\!\right)}{\sqrt{2 q_{ij[1]} + 2 |\mathbf{q}_{ij}|}}, \
- \frac{2 \left(\!
p_{ij[1]} q_{ij[2]} \!-\! p_{ij[2]} q_{ij[1]} \!-\! p_{ij[2]} |\mathbf{q}_{ij}|
\!\right)}{\sqrt{2 q_{ij[1]} + 2 |\mathbf{q}_{ij}|}}
\!\right) & \mbox{if} \ q_{ij[1]} \ge 0.
\end{array}
\right.
\label{p->P}
\end{eqnarray}
Substitution of transformations (\ref{Sec4.1:Levi-Civita}) and
(\ref{Sec4.1:Levi-Civita-p}) and their time differentiations into equation (7) yields
\begin{subnumcases}
{
}
\displaystyle
\mathbf{F}_{ij} = \mathbf{0}, \ 1 \le i < j \le N,
& \label{Sec4.1:Hamilton-eq-Q_{ij}} \\
\displaystyle
\mathbf{G}_{1j}
- \sum_{k=2}^{j-1} \mbox{\boldmath $\lambda$}_{kj}
+ \sum_{k=j+1}^N \mbox{\boldmath $\lambda$}_{jk}
= \mathbf{0}, \ 2 \le j \le N, & \label{Sec4.1:Hamilton-eq-P_{1j}} \\
\mathbf{G}_{ij}
+ \mbox{\boldmath $\lambda$}_{ij}
= \mathbf{0}, \ 1 < i < j \le N,
& \label{Sec4.1:Hamilton-eq-P_{ij}}
\end{subnumcases}
where
\begin{eqnarray}
&& \displaystyle
\mathbf{F}_{ij} \equiv
\frac{d}{dt} \mathbf{Q}_{ij} - \frac{m}{4 m_i m_j}
\frac{\mathbf{P}_{ij}}{|\mathbf{Q}_{ij}|^2}, \  1 \le i < j \le N, \nonumber\\
&& \displaystyle
\mathbf{G}_{ij} \equiv
\frac{1}{2 |\mathbf{Q}_{ij}|^2}
\!\left(\!
\frac{d}{dt} \mathbf{P}_{ij}
- \!\left(\! \frac{m}{\! 4 m_i m_j \!} |\mathbf{P}_{ij}|^2 - 2 m_i m_j \!\right)\!
\frac{\mathbf{Q}_{ij}}{|\mathbf{Q}_{ij}|^4} \!\right)\!
\mathbf{L}^{\top} (\mathbf{Q}_{ij}), \ 1 \le i < j \le N,
\label{Def-G} \\
&& \displaystyle \mbox{\boldmath $\lambda$}_{ij}
\equiv - \sum_{k=1}^{i-1} \frac{m_i m_j m_k}{m}
\left(
\frac{\mathbf{Q}_{ki}}{|\mathbf{Q}_{ki}|^6} \mathbf{L}^{\!\top\!}(\mathbf{Q}_{ki})
+ \frac{\mathbf{Q}_{ij}}{|\mathbf{Q}_{ij}|^6} \mathbf{L}^{\!\top\!}(\mathbf{Q}_{ij})
- \frac{\mathbf{Q}_{kj}}{|\mathbf{Q}_{kj}|^6} \mathbf{L}^{\!\top\!}(\mathbf{Q}_{kj})
\right) \nonumber\\
&& \qquad \quad \displaystyle
+ \sum_{k=i+1}^{j-1} \frac{m_i m_j m_k}{m}
\left(
\frac{\mathbf{Q}_{ik}}{|\mathbf{Q}_{ik}|^6} \mathbf{L}^{\!\top\!}(\mathbf{Q}_{ik})
+ \frac{\mathbf{Q}_{kj}}{|\mathbf{Q}_{kj}|^6} \mathbf{L}^{\!\top\!}(\mathbf{Q}_{kj})
- \frac{\mathbf{Q}_{ij}}{|\mathbf{Q}_{ij}|^6}
\mathbf{L}^{\!\top\!}(\mathbf{Q}_{ij}) \right)
\nonumber\\
&& \qquad \quad \displaystyle - \sum_{k=j+1}^N \frac{m_i m_j m_k}{m}
\left(
\frac{\mathbf{Q}_{ij}}{|\mathbf{Q}_{ij}|^6} \mathbf{L}^{\!\top\!}(\mathbf{Q}_{ij})
+ \frac{\mathbf{Q}_{jk}}{|\mathbf{Q}_{jk}|^6} \mathbf{L}^{\!\top\!}(\mathbf{Q}_{jk})
- \frac{\mathbf{Q}_{ik}}{|\mathbf{Q}_{ik}|^6} \mathbf{L}^{\!\top\!}(\mathbf{Q}_{ik})
\right), \quad 2 \le i < j \le N. \nonumber
\end{eqnarray}
In addition, through equation (\ref{Sec4.1:Levi-Civita}), the constraints in equation (\ref{Sec2.2:RC.Cst-q})
lead to
\begin{eqnarray}
\mathbf{\Phi}_{1jk} (\mathbf{Q})
\equiv \!\!\left[\!\!
\begin{array}{c}
\Phi_{1jk[1]} (\mathbf{Q}) \\
\Phi_{1jk[2]} (\mathbf{Q})
\end{array}
\!\!\right]^{\top}\!\!
\equiv \!\!\
\mathbf{Q}_{1j} \mathbf{L} (\mathbf{Q}_{1j})^{\top}
+ \mathbf{Q}_{jk} \mathbf{L} (\mathbf{Q}_{jk})^{\top}
- \mathbf{Q}_{1k} \mathbf{L} (\mathbf{Q}_{1k})^{\top}
= \mathbf{0},\ 2 \le j < k \le N,
\label{Sec4.1:Regularized-Constraint}
\end{eqnarray}
where
$\mathbf{Q} =
(\!\!
\begin{array}{c:c:c|c:c:c|c|c}
\mathbf{Q}_{1,2}, & \cdots & \mathbf{Q}_{1,N}, &
\mathbf{Q}_{2,3}, & \cdots & \mathbf{Q}_{2,N}, &
\cdots & \mathbf{Q}_{N-1,N}
\end{array}
\!\!)$ $\in \mathbb{R}^{N(N-1)}$
is a position vector.
The system composed of equations (15) and (\ref{Sec4.1:Regularized-Constraint})
represents the motion of the G$N$BP, which is described by the Levi-Civita
variables and the Lagrange multipliers $\mbox{\boldmath $\lambda$}_{jk}$.
This system is governed by the following Hamiltonian:
\begin{eqnarray}
H_{\mbox{\tiny LC}}
\equiv \sum_{i=1}^{N-1} \sum_{j=i+1}^N
\left( \frac{m}{8 m_i m_j} \frac{|\mathbf{P}_{ij}|^2}{|\mathbf{Q}_{ij}|^2}
- \frac{m_i m_j}{|\mathbf{Q}_{ij}|^2} \right)
+ \sum_{j=2}^{N-1} \sum_{k=j+1}^N \mbox{\boldmath $\Phi$}_{1jk}
(\mathbf{Q}) \cdot \mbox{\boldmath $\lambda$}_{jk},
\label{H_LC}
\end{eqnarray}
which is obtained by substituting equations (\ref{Sec4.1:Levi-Civita}) and
(\ref{Sec4.1:Levi-Civita-p}) into $H_{\mbox{\footnotesize rel}}$ defined
by equation (\ref{H_rel}).
\par
Using equation (\ref{Sec4.1:Hamilton-eq-P_{ij}}), we rewrite
equation (\ref{Sec4.1:Hamilton-eq-P_{1j}}) as the following identities:
\begin{eqnarray}
\displaystyle
\sum_{i=1}^{j-1} \mathbf{G}_{ij} - \sum_{i=j+1}^N \mathbf{G}_{ji}
= \mathbf{0}, \ 2 \le j \le N.
\label{Sec3.1:modified-Lagrange-type.Cn-Eq-p}
\end{eqnarray}
The new system composed of equations (\ref{Sec4.1:Hamilton-eq-Q_{ij}}),
(\ref{Sec4.1:Regularized-Constraint}), and
(\ref{Sec3.1:modified-Lagrange-type.Cn-Eq-p}) describes the same motion
as the system composed of equations (15) and (\ref{Sec4.1:Regularized-Constraint}).
The number of dependent variables, $P_{ij[1]}$, $P_{ij[2]}$, $Q_{ij[1]}$, and
$Q_{ij[2]}$ in the new system is $2N(N-1)$ rather than $4N$, which is the
number of dependent variables, $p'_{i[1]}$, $p'_{i[2]}$, $q'_{i[1]}$, and
$q'_{i[2]}$, in system (\ref{Sec2.1:CQ.diff2-q}).
Actually, as the total number of masses $N$ increases,
the number of equations in the new system increases more rapidly.
Therefore, the computational cost of the new system is very high for
large $N$.
We call this system the {\itshape redundant general three-body
problem (RG$N$BP)}.
\par
$H_{\mbox{\tiny LC}}$ is conserved by the RG$N$BP as
well as the system composed of equations (15) and (\ref{Sec4.1:Regularized-Constraint}).
Because of equation (\ref{Sec4.1:Regularized-Constraint}), the value of
$H_{\mbox{\tiny LC}}$ defined by equation (\ref{H_LC}) is equivalent to
\begin{eqnarray}
\displaystyle
h_{\mbox{\tiny LC}}
= \sum_{i=1}^{N-1} \sum_{j=i+1}^N \left(
\frac{m}{8 m_i m_j} \frac{|\mathbf{P}_{ij}|^2}{|\mathbf{Q}_{ij}|^2}
- \frac{m_i m_j}{|\mathbf{Q}_{ij}|}
\right).
\label{h_LC}
\end{eqnarray}
Further, $h_{\mbox{\tiny LC}}$ equals the value of $H$ defined by equation (\ref{Sec2.1:CQ.H})
because $h_{\mbox{\footnotesize rel}}$ defined by (\ref{h_rel}) is
transformed to $h_{\mbox{\tiny LC}}$ through equations (\ref{Sec4.1:Levi-Civita})
and (\ref{Sec4.1:Levi-Civita-p}), and $h_{\mbox{\footnotesize rel}}$ is
the value of $H$ as described in Section 2.2.1.
\subsection{Chain Regularization of General $N$-body Problem}
In this section, we rewrite the RG$N$BP using only
$\mathbf{P}_{k,k+1}$ and $\mathbf{Q}_{k,k+1}$ $(1 \le k \le N-1)$
so that we reduce the redundancy of the problem, which incurs a high
computational cost.
The vectors $\mathbf{Q}_{k,k+1}$ and $\mathbf{P}_{k,k+1}$ are related
to the chained position vectors $\mathbf{q}_{k,k+1}$ and their momenta
$\mathbf{p}_{k,k+1}$ conjugate to $\mathbf{q}_{k,k+1}$, respectively.
\par
First, we express the vectors $\mathbf{Q}_{ij}$ related to the
non-chained position vectors $\mathbf{q}_{ij}$ $(1 \le i < i+2 \le j \le
N)$ as functions of $\mathbf{Q}_{k,k+1}$ related to the
chained position vectors $\mathbf{q}_{k,k+1}$ $(1 \le k \le N-1)$.
The constraints in equation (\ref{Sec4.1:Regularized-Constraint}) restrict the possible
positions of the RG$N$BP to the $(N-1)-(N-1)(N-2)=2(N-1)$-dimensional configuration manifold
\begin{eqnarray}
\mathcal{Q} =
\left\{ \mathbf{Q} \in \mathbb{R}^{N(N-1)} \ | \
\mathbf{\Phi} (\mathbf{Q}) = \mathbf{0}_{1 \times (N-1)(N-2)} \right\},
\label{manifold_Q}
\end{eqnarray}
where
$\mathbf{\Phi} (\mathbf{Q}) =
(\!\!
\begin{array}{c:c:c|c:c:c|c|c}
\mathbf{\Phi}_{1,2,3} (\mathbf{Q}),
& \cdots &
\mathbf{\Phi}_{1,2,N} (\mathbf{Q}), &
\mathbf{\Phi}_{1,3,4} (\mathbf{Q}), &
\cdots & \mathbf{\Phi}_{1,3,N} (\mathbf{Q}), &
\cdots & \mathbf{\Phi}_{1,N-1,N} (\mathbf{Q})
\end{array}
\!\!)$ $\in \mathbb{R}^{(N-1)(N-2)}$
is a constraint function vector.
Accordingly, the set of $2(N-1)$ variables
$\left\{
\begin{array}{c:c:c:c}
Q_{1,2[1]}, Q_{1,2[2]}, & Q_{2,3[1]}, Q_{2,3[2]}, & \cdots &
Q_{N-1,N[1]}, Q_{N-1,N[2]}
\end{array} \right\}$, which corresponds to the $(N-1)$ vectors
$\mathbf{Q}_{k,k+1}$ $(1 \le k \le N-1)$, acts as a basis for
the manifold $\mathcal{Q}$.
Then, every vector $\mathbf{Q}_{ij} \in \mathbb{R}^{(N-1)(N-2)}$
$(1 \le i < i+2 \le j \le N)$ related to a non-chained position vector
$\mathbf{q}_{ij}$ fulfills the following relation:
\begin{eqnarray}
\mathbf{Q}_{ij} \mathbf{L} (\mathbf{Q}_{ij})^{\top}
= \sum_{k=i}^{j-1} \mathbf{Q}_{k,k+1} \mathbf{L}
(\mathbf{Q}_{k,k+1})^{\top}, \ 1 \le i < i+2 \le j \le N.
\label{Q_{ij}}
\end{eqnarray}
The solutions of equation (\ref{Q_{ij}}) for $\mathbf{Q}_{ij}$ are
\begin{eqnarray}
\displaystyle
\mathbf{Q}_{ij} = \tilde{\mathbf{Q}}_{ij}
\equiv
\left(
\displaystyle \sqrt{\frac{a_{ij} + b_{ij}}{2}}, \
\frac{\displaystyle \sqrt{2} \sum_{k=i}^{j-1} Q_{k,k+1[1]} Q_{k,k+1[2]}}
{\displaystyle \sqrt{a_{ij} + b_{ij}}}
\right), \quad 1 \le i < i+2 \le j \le N,
\label{tilde_Qij}
\end{eqnarray}
where
\begin{eqnarray*}
&& \displaystyle
a_{ij} \equiv \sqrt{
\left( \sum_{k=i}^{j-1} \left( Q_{k,k+1[1]}^2 - Q_{k,k+1[2]}^2 \right) \right)^2
+ 4 \left( \sum_{k=i}^{j-1} Q_{k,k+1[1]} Q_{k,k+1[2]} \right)^2
}, \nonumber\\
&& \displaystyle
b_{ij} \equiv
\sum_{k=i}^{j-1} \left( Q_{k,k+1[1]}^2 - Q_{k,k+1[2]}^2 \right),
\quad 1 \le i < i+2 \le j \le N.
\end{eqnarray*}
\par
Next, we write the vectors $\mathbf{P}_{ij}$ $(1 \le i < i+2 \le j \le
N)$ related to the momentum vectors $\mathbf{p}_{ij}$ as
functions of $\mathbf{Q}_{k,k+1}$ and $\mathbf{P}_{k,k+1}$ $(1 \le k \le
N-1)$ related to the momentum vectors $\mathbf{p}_{k,k+1}$.
The momenta of the RG$N$BP, $\mathbf{P}_{jk}$, satisfy
the following constraints:
\begin{eqnarray}
\displaystyle
\mathbf{\Psi}_{1jk}(\mathbf{P}, \mathbf{Q})
\!\equiv\!
\frac{1}{\! m_1 m_j \!} \frac{\mathbf{P}_{1j} \mathbf{L}
(\mathbf{Q}_{1j})^{\top}}{|\mathbf{Q}_{1j}|^2}
\!+\! \frac{1}{\! m_j m_k \!} \frac{\mathbf{P}_{jk} \mathbf{L}
(\mathbf{Q}_{jk})^{\top}}{|\mathbf{Q}_{jk}|^2}
\!-\! \frac{1}{\! m_1 m_k \!}  \frac{\mathbf{P}_{1k} \mathbf{L}
(\mathbf{Q}_{1k})^{\top}}{|\mathbf{Q}_{1k}|^2}
\!=\! \mathbf{0}, \ 2 \le j < k \le N,
\label{Refularized-constraint-P}
\end{eqnarray}
which is obtained by substituting equation (\ref{Sec4.1:Hamilton-eq-Q_{ij}})
into the time derivative of $\mathbf{\Phi}_{1jk} (\mathbf{Q})$.
Here,
$\mathbf{P} =
(\!\!
\begin{array}{c:c:c|c:c:c|c|c}
\mathbf{P}_{1,2}, & \cdots & \mathbf{P}_{1,N}, &
\mathbf{P}_{2,3}, & \cdots & \mathbf{P}_{2,N}, &
\cdots & \mathbf{P}_{N-1,N}
\end{array}
\!\!)$ $\in \mathbb{R}^{N(N-1)}$ is a momentum vector.
The constraints in equation (\ref{Refularized-constraint-P}) restrict the possible
momenta of the RG$N$BP to the $N(N-1)-(N-1)(N-2)=2(N-1)$-dimensional manifold
\begin{eqnarray*}
\mathcal{P} = \left\{
\mathbf{P} \in \mathbb{R}^{N(N-1)} \ | \
\mathbf{\Psi} (\mathbf{P}, \mathbf{Q}) = \mathbf{0}_{1 \times (N-1) (N-2)}
\right\},
\end{eqnarray*}
where
{\footnotesize
$\mathbf{\Psi} (\mathbf{P}, \mathbf{Q}) =
(\!\!
\begin{array}{c:c:c|c:c:c|c|c}
\mathbf{\Psi}_{1,2,3} (\mathbf{P}, \mathbf{Q}),
& \cdots &
\mathbf{\Psi}_{1,2,N} (\mathbf{P}, \mathbf{Q}), &
\mathbf{\Psi}_{1,3,4} (\mathbf{P}, \mathbf{Q}), &
\cdots & \mathbf{\Psi}_{1,3,N} (\mathbf{P}, \mathbf{Q}), &
\cdots & \mathbf{\Psi}_{1,N-1,N} (\mathbf{P}, \mathbf{Q})
\end{array}
\!\!)$}
$\in \mathbb{R}^{(N-1)(N-2)}$
is a constraint function vector.
Thus, the $(N-1)$ vectors $\mathbf{P}_{k,k+1}$ $(1 \le k \le N-1)$
are a basis for the manifold $\mathcal{P}$.
Each vector $\mathbf{P}_{ij} \in \mathbb{R}^{(N-1)(N-2)}$
$(1 \le i < i+2 \le j \le N)$ related to a non-chained momentum vector
$\mathbf{p}_{ij}$ can be uniquely expressed as follows:
\begin{eqnarray}
\displaystyle
\frac{\mathbf{P}_{ij} \mathbf{L} (\mathbf{Q}_{ij})^{\top}}{|\mathbf{Q}_{ij}|^2}
\!=\! m_i m_j \sum_{k=i}^{j-1} \frac{1}{m_k m_{k+1}}
\frac{\mathbf{P}_{k,k+1} \mathbf{L} (\mathbf{Q}_{k,k+1})^{\top}}
{|\mathbf{Q}_{k,k+1}|^2},
\ 1 \le i < i+2 \le j \le N. \label{P_{ij}}
\end{eqnarray}
We solve equation (\ref{P_{ij}}) for $\mathbf{P}_{ij}$ $(1 \le i < i+2 \le j \le
N)$ and subsequently substitute equation (\ref{tilde_Qij}) into the resulting
solution.
Then, we obtain
{\footnotesize
\begin{eqnarray}
&& \hspace*{-10mm} \mathbf{P}_{ij} = \tilde{\mathbf{P}}_{ij} \nonumber\\
&& \hspace*{-10mm} \displaystyle \equiv m_i m_j
\!\!\left[\!\!
\begin{array}{c}
\displaystyle
\tilde{Q}_{i,j[1]} \sum_{k=i}^{j-1}
\frac{P_{k,k+1[1]} Q_{k,k+1[1]} \!-\! P_{k,k+1[2]} Q_{k,k+1[2]}}
{m_k m_{k+1} |\mathbf{Q}_{k,k+1}|^2}
+ \tilde{Q}_{i,j[2]} \sum_{k=i}^{j-1}
\frac{P_{k,k+1[1]} Q_{k,k+1[2]} \!+\! P_{k,k+1[2]} Q_{k,k+1[1]}}
{m_k m_{k+1} |\mathbf{Q}_{k,k+1}|^2} \\
\displaystyle
\tilde{Q}_{i,j[1]} \sum_{k=i}^{j-1}
\frac{P_{k,k+1[1]} Q_{k,k+1[2]} \!+\! P_{k,k+1[2]} Q_{k,k+1[1]}}
{m_k m_{k+1} |\mathbf{Q}_{k,k+1}|^2}
- \tilde{Q}_{i,j[2]} \sum_{k=i}^{j-1}
\frac{P_{k,k+1[1]} Q_{k,k+1[1]} \!-\! P_{k,k+1[2]} Q_{k,k+1[2]}}
{m_k m_{k+1} |\mathbf{Q}_{k,k+1}|^2}
\end{array}
\!\!\right]^{\top}\!\!, \nonumber\\
&& \hspace*{105mm} 1 \le i < i+2 \le j \le N.
\label{tilde_Pij}
\end{eqnarray}
}
\par
Using equations (\ref{tilde_Qij}) and (\ref{tilde_Pij}), we can rewrite
equation (\ref{Sec3.1:modified-Lagrange-type.Cn-Eq-p}) as
\begin{eqnarray}
&& \displaystyle
\sum_{i=1}^{j-2} \tilde{\mathbf{G}}_{ij} - \sum_{i=j+2}^N
\tilde{\mathbf{G}}_{ji}
= - \mathbf{G}_{j-1,j} + \mathbf{G}_{j,j+1}, \ 2 \le j \le N-1,
\nonumber\\
&& \displaystyle
\sum_{i=1}^{N-2} \tilde{\mathbf{G}}_{i,N} = - \mathbf{G}_{N-1,N},
\label{Relation-tilde_G}
\end{eqnarray}
where
\begin{eqnarray*}
\displaystyle
\tilde{\mathbf{G}}_{ij} \equiv
\frac{1}{2 |\tilde{\mathbf{Q}}_{ij}|^2}
\!\left(\!
\frac{d}{dt} \tilde{\mathbf{P}}_{ij}
- \!\left(\! \frac{m}{\! 4 m_i m_j \!} |\tilde{\mathbf{P}}_{ij}|^2
- 2 m_i m_j \!\right)\!
\frac{\tilde{\mathbf{Q}}_{ij}}{|\tilde{\mathbf{Q}}_{ij}|^4} \!\right)\!
\mathbf{L}^{\top} (\tilde{\mathbf{Q}}_{ij}), \ 1 \le i < i+2 \le j \le N,
\end{eqnarray*}
and $\mathbf{G}_{j-1,j}$, $\mathbf{G}_{j,j+1}$, and $\mathbf{G}_{N-1,N}$
are defined by equation (\ref{Def-G}).
Note that equation (\ref{Relation-tilde_G}) is described by only $4(N-1)$
variables, $Q_{k,k+1[1]}$, $Q_{k,k+1[2]}$, $P_{k,k+1[1]}$, and
$P_{k,k+1[2]}$, related to the chained vectors $\mathbf{q}_{k,k+1}$ and
$\mathbf{p}_{k,k+1}$ $(1 \le k \le N-1)$.
In addition, the $2(N-1)$ vectors $\mathbf{Q}_{k,k+1}$ and $\mathbf{P}_{k,k+1}$
$(1 \le k \le N-1)$ satisfy equation (\ref{Sec4.1:Hamilton-eq-Q_{ij}}); namely,
\begin{eqnarray}
\displaystyle
\frac{d}{dt} \mathbf{Q}_{k,k+1} = \frac{m}{4 m_k m_{k+1}}
\frac{\mathbf{P}_{k,k+1}}{|\mathbf{Q}_{k,k+1}|^2},\quad
1 \le k \le N-1. \label{Sec4.1:Hamilton-eq-Q_{i,i+1}}
\end{eqnarray}
We call the system composed of equations (\ref{Relation-tilde_G}) and
(\ref{Sec4.1:Hamilton-eq-Q_{i,i+1}})
the {\itshape chain regularization of G$N$BP (CRG$N$BP)}.
We clarify that the CRG$N$BP describes the same motion as the RG$N$BP composed of
equations (\ref{Sec4.1:Hamilton-eq-Q_{ij}}), (\ref{Sec4.1:Regularized-Constraint}), and
(\ref{Sec3.1:modified-Lagrange-type.Cn-Eq-p}) in the following lemma.
\newtheorem{lmm}{Lemma}
\begin{lmm}
Suppose
\begin{enumerate}
\item[(i)]
The $2(N-1)$ vectors $\mathbf{Q}_{k,k+1}$ and $\mathbf{P}_{k,k+1}$
$(1 \le k \le N-1)$ are the solutions of the CRG$N$BP composed of
equations (\ref{Relation-tilde_G}) and (\ref{Sec4.1:Hamilton-eq-Q_{i,i+1}}).
\item[(ii)]
The $(N-1)(N-2)$ vectors $\mathbf{Q}_{ij}$ and $\mathbf{P}_{ij}$
$(1 \le i < i+2 \le j \le N)$ are given by equations (\ref{tilde_Qij})
and (\ref{tilde_Pij}).
\end{enumerate}
Then, the $N(N-1)$ vectors $\mathbf{Q}_{ij}$ and $\mathbf{P}_{ij}$
$(1 \le i < j \le N)$ satisfy the RG$N$BP composed of
equations (\ref{Sec4.1:Hamilton-eq-Q_{ij}}), (\ref{Sec4.1:Regularized-Constraint}),
and (\ref{Sec3.1:modified-Lagrange-type.Cn-Eq-p}).
\end{lmm}
{\itshape Proof.} \ \\
(a)\ {\itshape Derivation of equation (\ref{Sec4.1:Hamilton-eq-Q_{ij}})} \ \\
The vectors $\mathbf{Q}_{ij}$ and $\mathbf{P}_{ij}$ $(1 \le i < i+2 \le
j \le N)$ defined by equations (\ref{tilde_Qij}) and (\ref{tilde_Pij}) satisfy
(\ref{P_{ij}}).
In addition, using equation (\ref{Sec4.1:Hamilton-eq-Q_{i,i+1}}), equation (\ref{P_{ij}}) is
rewritten as
\begin{eqnarray}
\displaystyle
\frac{m}{4 m_1 m_j}
\frac{\mathbf{P}_{ij} \mathbf{L} (\mathbf{Q}_{ij})^{\top}}{|\mathbf{Q}_{ij}|^2}
= \sum_{k=i}^{j-1}
\frac{d}{dt} \mathbf{Q}_{k,k+1} \mathbf{L} (\mathbf{Q}_{k,k+1})^{\top}, \
1 \le i < i+2 \le j \le N. \label{(a-1)}
\end{eqnarray}
Similarly, the vectors $\mathbf{Q}_{ij}$ $(1 \le i < i+2 \le j \le N)$
in equation (\ref{tilde_Qij}) satisfy equation (\ref{Q_{ij}}), so the time derivative of
(\ref{Q_{ij}}) is also fulfilled.
The derivative is written in the following form:
\begin{eqnarray}
\displaystyle
\frac{d}{dt} \mathbf{Q}_{ij} \mathbf{L} (\mathbf{Q}_{ij})^{\top}
= \sum_{k=i}^{j-1} \frac{d}{dt} \mathbf{Q}_{k,k+1} \mathbf{L}
(\mathbf{Q}_{k,k+1})^{\top}, \ 1 \le i < i+2 \le j \le N.
\label{(a-2)}
\end{eqnarray}
Because the r.h.s. of equation (\ref{(a-1)}) coincides with the r.h.s. of
equation (\ref{(a-2)}), the left-hand sides of equations (\ref{(a-1)}) and
(\ref{(a-2)}) are equal.
Accordingly, equation (\ref{Sec4.1:Hamilton-eq-Q_{ij}}) in the CRG$N$BP is obtained.
\ \\
\ \\
(b)\ {\itshape Derivation of equation (\ref{Sec4.1:Regularized-Constraint})}
\ \\
We have already shown that equation (\ref{tilde_Qij}) is equivalent to
equation (\ref{Q_{ij}}).
In addition, substituting equation (\ref{Q_{ij}}) into $\mathbf{\Phi}_{1jk} (\mathbf{Q})$
in equation (\ref{Sec4.1:Regularized-Constraint}) yields $\mathbf{0}$.
Accordingly, the vectors $\mathbf{Q}_{ij}$ $(1 \le i < i+2 \le j \le N)$ in
equation (\ref{tilde_Qij}) satisfy equation (\ref{Sec4.1:Regularized-Constraint}).
\ \\
\ \\
(c)\ {\itshape Derivation of equation (\ref{Sec3.1:modified-Lagrange-type.Cn-Eq-p})}
\ \\
Substituting equations (\ref{tilde_Qij}) and (\ref{tilde_Pij}) into
equation (\ref{Sec3.1:modified-Lagrange-type.Cn-Eq-p}) yields equation
(\ref{Relation-tilde_G}).
Namely, equation (\ref{Relation-tilde_G}) is equal to
equation (\ref{Sec3.1:modified-Lagrange-type.Cn-Eq-p}) under the conditions
in equations (\ref{tilde_Qij}) and (\ref{tilde_Pij}).
Thus, the $N(N-1)$ vectors $\mathbf{Q}_{ij}$ and $\mathbf{P}_{ij}$
$(1 \le i < j \le N)$ satisfy equation (\ref{Sec3.1:modified-Lagrange-type.Cn-Eq-p}).
\qquad \fbox{}
\par
For large $N$, the number of dependent variables $P_{i,i+1[1]}$,
$P_{i,i+1[2]}$, $Q_{i,i+1[1]}$, and $Q_{i,i+1[2]}$ in the CRG$N$BP,
$4(N-1)$, is remarkably smaller than $2N(N-1)$, which is the number of
dependent variables in the RG$N$BP composed of
equations (\ref{Sec4.1:Hamilton-eq-Q_{ij}}), (\ref{Sec4.1:Regularized-Constraint}),
and (\ref{Sec3.1:modified-Lagrange-type.Cn-Eq-p}).
Therefore, the computational cost of the CRG$N$BP is much lower than
that of the RG$N$BP.
\section{Energy-momentum Integrator for General $N$-body Problem}
We use the d'Alembert-type scheme \citep{Betsch2005} to
discretize the G$N$BP, which leads to the G$3$BP for $N = 3$.
Regardless of the number of masses $N$, the forms of
equation (\ref{Sec4.1:Hamilton-eq-Q_{ij}}) and the functions $\mathbf{G}_{ij}$
defined by equation (\ref{Def-G}) are invariant for each $(i, j)$.
\par
In Section 3.1, we discretize the RG$N$BP composed of
equations (\ref{Sec4.1:Hamilton-eq-Q_{ij}}), (\ref{Sec4.1:Regularized-Constraint}),
and (\ref{Sec3.1:modified-Lagrange-type.Cn-Eq-p})
so that, for $N=3$, the discretized forms of
equation (\ref{Sec4.1:Hamilton-eq-Q_{ij}}) and $\mathbf{G}_{i,j}$ coincide with
the counterparts of the d'Alembert-type scheme for the G$3$BP
\citep{Minesaki-2013a}.
However, because the discrete-time problem including these forms has redundant
dependent variables, it suffers from high computational cost for a large
number of masses $N$.
In Section 3.2, we remove the redundancy using some constraints, so
we obtain the discretization of the CRG$N$BP composed of
equations (\ref{Relation-tilde_G}) and (\ref{Sec4.1:Hamilton-eq-Q_{i,i+1}}).
The resulting discrete-time problem preserves all the conserved quantities
except for the angular momentum precisely.
\subsection{Discrete-time General $N$-body Problem with Redundant Variables}
For $N=3$, the RG$N$BP composed of
equations (\ref{Sec4.1:Hamilton-eq-Q_{ij}}), (\ref{Sec4.1:Regularized-Constraint}),
and (\ref{Sec3.1:modified-Lagrange-type.Cn-Eq-p}) represents the
regularized G$3$BP:
\begin{eqnarray}
\left\{
\begin{array}{l}
\displaystyle \frac{d}{dt} \mathbf{Q}_{ij}
= \frac{m}{4 m_i m_j} \frac{\mathbf{P}_{ij}}{|\mathbf{Q}_{ij}|^2},
\ 1 \le i < j \le 3, \\
\displaystyle
\mathbf{G}_{12} - \mathbf{G}_{23} = \mathbf{0}, \quad
\mathbf{G}_{13} + \mathbf{G}_{23} = \mathbf{0}, \\
\mathbf{\Phi}_{123} (\mathbf{Q}) = \mathbf{0},
\end{array}
\right. \label{d-G3BP}
\end{eqnarray}
where $\mathbf{G}_{ij}$ is defined by equation (\ref{Def-G}).
Applying the extension of the d'Alembert-type scheme in \citep{Betsch2005},
we gave the d-G$3$BP \citep{Minesaki-2013a} with a time step $\Delta t =
t^{(n+1)} - t^{(n)}$, namely, the discretization of system
(\ref{d-G3BP})
\footnote{
This discretization coincides with discrete-time system (43) in
\citep{Minesaki-2013a}.
Note that the vectors $\left(\mathbf{\bullet} \right)_{12}$,
$\left(\mathbf{\bullet}\right)_{23}$,
and $\left(\mathbf{\bullet}\right)_{13}$ correspond to
$\left(\mathbf{\bullet}\right)_{3}$, $\left(\mathbf{\bullet}\right)_{1}$,
and $- \left(\mathbf{\bullet}\right)_{2}$
in \citep{Minesaki-2013a}, respectively.
}:
\begin{eqnarray}
\left\{
\begin{array}{l}
\mathbf{F}_{ij}^{(n+1)} = \mathbf{0}, \ 1 \le i < j \le 3, \\
\displaystyle
\mathbf{G}_{12}^{(n+1)} - \mathbf{G}_{23}^{(n+1)} = \mathbf{0}, \quad
\mathbf{G}_{13}^{(n+1)} + \mathbf{G}_{23}^{(n+1)} = \mathbf{0}, \\
\mathbf{\Phi}_{123} (\mathbf{Q}^{(n+1)}) = \mathbf{0},
\end{array}
\right. \label{(A)}
\end{eqnarray}
where $\mathbf{Q}_{ij}^{(l)} = ( Q_{ij[1]}^{(l)}, Q_{ij[2]}^{(l)} )$
and $\mathbf{P}_{ij}^{(l)} = ( P_{ij[1]}^{(l)}, P_{ij[2]}^{(l)} )$
at time $t^{(l)}$ $(l=n,n+1)$ for $1 \le i < j \le 3$; we define the
midpoint value $(\bullet)^{(n+1/2)} \equiv \left[ (\bullet)^{(n+1)} +
(\bullet)^{(n)} \right] /2$ of the function $(\bullet) (t)$, and
{\footnotesize
\begin{eqnarray}
&& \hspace*{-5mm} \displaystyle
\mathbf{F}_{ij}^{\!(n+1)\!}
\equiv \frac{\mathbf{Q}_{ij}^{(n+1)} - \mathbf{Q}_{ij}^{(n)}}{\Delta t}
- \frac{m}{8 m_i m_j}
\frac{|\mathbf{Q}_{ij}^{(n+1)}|^2 +
|\mathbf{Q}_{ij}^{(n)}|^2}{|\mathbf{Q}_{ij}^{(n+1)}|^2
|\mathbf{Q}_{ij}^{(n)}|^2} \mathbf{P}_{ij}^{(n+1/2)}, \
1 \le i < j \le 3, \label{def-dFij} \\
&& \hspace*{-5mm} \mathbf{G}_{ij}^{\!(n+1)} \nonumber\\
&& \hspace*{-5mm} \equiv \displaystyle
\frac{1}
{2 |\mathbf{Q}_{ij}^{\!(n \!+\! 1/2)\!}|^2}
\!\left(\!
\frac{\mathbf{P}_{ij}^{\!(n \!+\! 1)\!} \!-\! \mathbf{P}_{ij}^{(n)}}
{\Delta t} \!-\! \frac{1}{|\mathbf{Q}_{ij}^{\!(n \!+\! 1)\!}|^2
|\mathbf{Q}_{ij}^{(n)}|^2}
\!\left(\!
\frac{m}{8 m_i m_j}
\!\left(\! |\mathbf{P}_{ij}^{\!(n \!+\! 1)\!}|^2 \!+\!
   |\mathbf{P}_{ij}^{(n)}|^2 \!\right)\!
\!-\! 2 m_i m_j
\!\right)\! \mathbf{Q}_{ij}^{(n \!+\! 1/2)}
\!\right)\! \mathbf{L} \left(\! \mathbf{Q}_{ij}^{\!(n+1/2)\!}
		       \!\right)^{\top}, \nonumber\\
&& \hspace*{120mm} 1 \le i < j \le 3.  \label{def-dGij}
\end{eqnarray}
}
\par
For $N \ge 3$, the RG$N$BP composed of
equations (\ref{Sec4.1:Hamilton-eq-Q_{ij}}), (\ref{Sec4.1:Regularized-Constraint}),
and (\ref{Sec3.1:modified-Lagrange-type.Cn-Eq-p})
involves the same $\mathbf{F}_{ij}$ and $\mathbf{G}_{ij}$ as
the d-G$3$BP in equation (\ref{(A)}) outside the range of $i$ and $j$.
Therefore, ignoring this range, we adopt $\mathbf{F}_{ij}^{(n+1)}$ and
$\mathbf{G}_{ij}^{(n+1)}$ defined by equations (\ref{def-dFij}) and
(\ref{def-dGij}) as the discrete analogs of $\mathbf{F}_{ij}$ and
$\mathbf{G}_{ij}$.
Concretely, the discrete-time system, which approximately describes
the motion of the RG$N$BP in a typical time interval $I^{(n)} =
[t^{(n)}, t^{(n+1)}]$ with a corresponding time step
$\Delta t = t^{(n+1)} - t^{(n)}$, is given as follows:
\begin{subnumcases}
{}
\displaystyle \mathbf{F}_{ij}^{(n+1)} = \mathbf{0}, \quad
1 \le i < j \le N, & \label{dQ-redundant-GNBP} \\
\displaystyle
\sum_{i=1}^{j-1} \mathbf{G}_{ij}^{(n+1)} - \sum_{i=j+1}^N
\mathbf{G}_{ji}^{(n+1)} = \mathbf{0}, \ 2 \le j \le N, &
\label{dP-redundant-GNBP} \\
\displaystyle
\mbox{$\mathbf{\Phi}$}_{1ij} (\mathbf{Q}^{(n+1)}) = \mathbf{0}, \
2 \le i < j \le N. & \label{dPhi-redundant-GNBP}
\end{subnumcases}
$\mathbf{Q}^{(n)} \in \mathcal{Q}$ and $\mathbf{P}^{(n)} \in \mathbb{R}^{N(N-1)}$
are given quantities at time node $t^{(n)}$, where
$\mathcal{Q}$ was already defined by (\ref{manifold_Q}).
In the following, we call this system the
{\itshape discrete-time redundant general $N$-body problem (d-RG$N$BP)}.
It can be used to calculate the unknown $\mathbf{Q}^{(n+1)}$ and
$\mathbf{P}^{(n+1)} =
(\!\!
\begin{array}{c:c:c|c:c:c|c|c}
\mathbf{P}_{1,2}^{(n+1)}, & \cdots & \mathbf{P}_{1,N}^{(n+1)}, &
\mathbf{P}_{2,3}^{(n+1)}, & \cdots & \mathbf{P}_{2,N}^{(n+1)}, &
\cdots & \mathbf{P}_{N-1,N}^{(n+1)}
\end{array}
\!\!)$ $\in \mathbb{R}^{N(N-1)}$
Note that $\mathbf{Q}^{(n+1)} \in \mathcal{Q}$ because of equation
(\ref{dPhi-redundant-GNBP}).
\par
The d-RG$N$BP preserves all the conserved quantities, except the angular momentum,
precisely.
This is shown in the following theorem.
\newtheorem{thm}{Theorem}
\begin{thm} (Conserved quantities of d-RG$N$BP) \
The d-RG$N$BP (35) keeps the following three
conserved quantities exactly:
\begin{enumerate}
\item[1.] the Hamiltonian defined by (\ref{Sec2.1:CQ.H}),
\item[2.]  the linear momentum $\displaystyle \mathbf{l} \equiv \sum_{i=1}^N
	  \mathbf{p}_i' = \mathbf{0}$,
\item[3.] the position of the center of mass $\displaystyle \mathbf{c}
	  \equiv \sum_{i=1}^N m_i \mathbf{q}_i' = \mathbf{0}$.
\end{enumerate}
\end{thm}
{\itshape Proof.}
\begin{enumerate}
\item[1.]
By using the multipliers $\mbox{\boldmath $\Lambda$}_{ij} = \left(
\Lambda_{ij[1]}, \Lambda_{ij[2]} \right)$ $(2 \le i < j \le
N)$, the d-RG$N$BP (35) is rewritten as
\begin{subnumcases}
{
}
\displaystyle \mathbf{F}_{ij}^{(n+1)} = \mathbf{0}, \quad 1 \le i < j
\le N, & \label{dQ_with_lambda} \\
\displaystyle \mathbf{G}_{1j}^{(n+1)}
- \sum_{k=2}^{j-1} \mbox{\boldmath $\Lambda$}_{kj}
+ \sum_{k=j+1}^N \mbox{\boldmath $\Lambda$}_{jk}
= \mathbf{0}, \quad j=2,\cdots,N, & \label{dP1j_with_lambda} \\
\displaystyle \mathbf{G}_{ij}^{(n+1)} + \mbox{\boldmath $\Lambda$}_{ij}
= \mathbf{0}, \quad
2 \le i < j \le N. & \label{dPij_with_lambda}
\end{subnumcases}
By using equation (\ref{dPij_with_lambda}), equation (\ref{dP1j_with_lambda}) is
expressed as equation (\ref{dP-redundant-GNBP}).
\par
We set the following functions:
{\footnotesize
\begin{eqnarray*}
&& \displaystyle
H_{1j} (\mathbf{P}_{1j}, \mathbf{Q}_{1j}) \equiv
\frac{m}{8 m_1 m_j} \frac{|\mathbf{P}_{1j}|^2}{|\mathbf{Q}_{1j}|^2}
- \frac{m_1 m_j}{|\mathbf{Q}_{1j}|^2}
+ \mathbf{Q}_{1j} \mathbf{L} (\mathbf{Q}_{1j})^{\top} \cdot
\left(
\sum_{i=2}^{j-1} \mbox{\boldmath $\Lambda$}_{ij}
- \sum_{i=j+1}^N \mbox{\boldmath $\Lambda$}_{ji}
\right), \ 2 \le j \le N, \\
&& \displaystyle
H_{ij}  (\mathbf{P}_{ij}, \mathbf{Q}_{ij}) \equiv
\displaystyle
\frac{m}{8 m_i m_j} \frac{|\mathbf{P}_{ij}|^2}{|\mathbf{Q}_{ij}|^2}
- \frac{m_i m_j}{|\mathbf{Q}_{ij}|^2}
+ \mathbf{Q}_{ij} \mathbf{L} (\mathbf{Q}_{ij})^{\top} \cdot
\mbox{\boldmath $\Lambda$}_{ij}, \quad 2 \le i < j \le N.
\end{eqnarray*}
}
Because (\ref{Sec4.1:Regularized-Constraint}) and
 \begin{eqnarray*}
\displaystyle
\sum_{i=1}^{N-1} \sum_{j=i+1}^N H_{ij} (\mathbf{P}_{ij}, \mathbf{Q}_{ij})
= \sum_{i=1}^{N-1} \sum_{j=i+1}^N
\left( \frac{m}{8 m_i m_j} \frac{|\mathbf{P}_{ij}|^2}{|\mathbf{Q}_{ij}|^2}
- \frac{m_i m_j}{|\mathbf{Q}_{ij}|^2} \right)
+ \sum_{j=2}^{N-1} \sum_{k=j+1}^N \mbox{\boldmath $\Phi$}_{1jk}
(\mathbf{Q}) \cdot \mbox{\boldmath $\Lambda$}_{jk},
\end{eqnarray*}
the value of the r.h.s. coincides with
$h_{\mbox{\tiny LC}}$ in equation (\ref{h_LC}),
which equals the value of $H_{\mbox{\tiny LC}}$ defined by equation (\ref{H_LC}).
For $2 \le j \le N$, scalar multiplication of equation (\ref{dQ_with_lambda}) for
$i=1$ and equation (\ref{dP1j_with_lambda}) by $\mathbf{P}_{1j}^{(n+1)} -
\mathbf{P}_{1j}^{(n)}$ and $- (\mathbf{Q}_{1j}^{(n+1)} -
\mathbf{Q}_{1j}^{(n)})$, respectively, and subsequent addition of the
two equations yield
\begin{eqnarray}
&& \hspace*{-10mm} \displaystyle
\frac{m}{8 m_1 m_j} \frac{|\mathbf{P}_{1j}^{(n+1)}|^2}{|\mathbf{Q}_{1j}^{(n+1)}|^2}
-
\frac{m}{8 m_1 m_j}
\frac{|\mathbf{P}_{1j}^{(n)}|^2}{|\mathbf{Q}_{1j}^{(n)}|^2}
- \left(
\frac{m_1 m_j}{|\mathbf{Q}_{1j}^{(n+1)}|^2}
- \frac{m_1 m_j}{|\mathbf{Q}_{1j}^{(n)}|^2}
\right) \nonumber\\
&& \hspace*{-10mm} \displaystyle
+ \left(
\mathbf{Q}_{1j}^{(n+1)} \mathbf{L} (\mathbf{Q}_{1j}^{(n+1)})^{\top}
- \mathbf{Q}_{1j}^{(n)} \mathbf{L} (\mathbf{Q}_{1j}^{(n)})^{\top}
\right) \cdot
\left( \sum_{i=2}^{j-1} \mathbf{\Lambda}_{ij}
- \sum_{i=j+1}^N \mathbf{\Lambda}_{ji}
\right) = \mathbf{0}, \ 2 \le j \le N.
\label{Conservation-H1j}
\end{eqnarray}
Equation (\ref{Conservation-H1j}) represents the conservation of
$H_{1j} (\mathbf{P}_{1j}, \mathbf{Q}_{1j})$
because it equals
$H_{1j} (\mathbf{P}_{1j}^{(n+1)}, \mathbf{Q}_{1j}^{(n+1)}) -
H_{1j} (\mathbf{P}_{1j}^{(n)}, \mathbf{Q}_{1j}^{(n)}) = 0$.
In addition, for $1 < i < j \le N$, scalar multiplication of
equation (\ref{dQ_with_lambda}) for $i \ge 2$ and equation (\ref{dPij_with_lambda}) by
$\mathbf{P}_{ij}^{(n+1)} - \mathbf{P}_{ij}^{(n)}$ and
$- (\mathbf{Q}_{ij}^{(n+1)} - \mathbf{Q}_{ij}^{(n)})$, respectively, and
subsequent addition of the two equations lead to
\begin{eqnarray}
&& \hspace*{-5mm} \displaystyle
\frac{m}{8 m_i m_j} \frac{|\mathbf{P}_{ij}^{(n+1)}|^2}{|\mathbf{Q}_{ij}^{(n+1)}|^2}
-
\frac{m}{8 m_i m_j}
\frac{|\mathbf{P}_{ij}^{(n)}|^2}{|\mathbf{Q}_{ij}^{(n)}|^2}
- \left(
\frac{m_i m_j}{|\mathbf{Q}_{ij}^{(n+1)}|^2}
- \frac{m_i m_j}{|\mathbf{Q}_{ij}^{(n)}|^2}
\right) \nonumber\\
&& \hspace*{-5mm} \displaystyle +
\left(
\mathbf{Q}_{ij}^{(n+1)} \mathbf{L} (\mathbf{Q}_{ij}^{(n+1)})^{\top}
- \mathbf{Q}_{ij}^{(n)} \mathbf{L} (\mathbf{Q}_{ij}^{(n)})^{\top}
\right) \cdot \mathbf{\Lambda}_{ij} = \mathbf{0}, \ 1 < i < j \le N.
\label{Conservation-Hij}
\end{eqnarray}
Equation (\ref{Conservation-Hij}) describes the conservation of
$H_{ij} (\mathbf{P}_{ij}, \mathbf{Q}_{ij})$
because it is equivalent to
$H_{ij} (\mathbf{P}_{ij}^{(n+1)}, \mathbf{Q}_{ij}^{(n+1)}) -
H_{ij} (\mathbf{P}_{ij}^{(n)}, \mathbf{Q}_{ij}^{(n)}) = 0$.
Accordingly, the d-RG$N$BP (35) preserves the value of
$h_{\mbox{\tiny LC}}$ defined by equation (\ref{h_LC}).
The d-RG$N$BP (35) conserves the value of $H$ defined by equation (\ref{Sec2.1:CQ.H})
because the value $h_{\mbox{\tiny LC}}$ is that of $H_{\mbox{\tiny LC}}$ in
equation (\ref{H_LC}), and the value of $h_{\mbox{\tiny LC}}$ equals that of $H$ in equation (1), as
	 described in Section 2.2.2.
\item[2].
The linear momentum $\mathbf{l}^{(n+1)}$ in the barycentric frame at
time node $t^{(n+1)}$ is given by
\begin{eqnarray}
\mathbf{l}^{(n+1)} = \sum_{i=1}^N \mathbf{p}_i^{'(n+1)}.
\label{Sec5.1:D-l}
\end{eqnarray}
Substituting equation (\ref{Sec2.2:qp_i}) into equation (\ref{Sec5.1:D-l}), we see
\begin{eqnarray*}
\mathbf{l}^{(n+1)} = \mathbf{0}.
\end{eqnarray*}
Therefore, the d-G$N$BP (35) preserves the linear
momentum $\mathbf{l}$, which is identically zero.
\item[3.]
The position of the center of mass $\mathbf{c}^{(n+1)}$ in the
barycentric frame at time node $t^{(n+1)}$ is expressed by
\begin{eqnarray}
\mathbf{c}^{(n+1)} = \sum_{i=1}^N m_i \mathbf{q}_i^{'(n+1)}.
\label{Sec5.1:D-c}
\end{eqnarray}
We substitute equation (\ref{Sec2.2:qp_i}) into equation (\ref{Sec5.1:D-c}).
Then, we can check
\begin{eqnarray*}
\mathbf{c}^{(n+1)} = \mathbf{0}.
\end{eqnarray*}
\end{enumerate}
As a result, the d-RG$N$BP (35) keeps the vector
value of the position of the center of motion $\mathbf{c}$ at
zero. \qquad \fbox{}
\par
For large $N$, the number of variables $P_{ij[1]}^{(n+1)}$,
$P_{ij[2]}^{(n+1)}$, $Q_{ij[1]}^{(n+1)}$, and $Q_{ij[1]}^{(n+1)}$ in the
d-RG$N$BP, $2N(N - 1)$, is remarkably larger than $4N$, which is
the number of dependent variables $p'_{i[1]}$, $p'_{i[2]}$,
$q'_{i[1]}$, and $q'_{i[2]}$ in the G$N$BP (\ref{Sec2.1:CQ.diff2-q}).
Thus, the computational cost of the d-RG$N$BP is very high.
\subsection{Discrete-time General $N$-body Problem with Chain Variables}
In this section, to reduce the high computational cost of the d-RG$N$BP, we
rewrite the d-RG$N$BP using only $\mathbf{Q}_{k,k+1}^{(n)}$,
$\mathbf{Q}_{k,k+1}^{(n+1)}$, $\mathbf{P}_{k,k+1}^{(n)}$, and
$\mathbf{P}_{k,k+1}^{(n+1)}$ $(1 \le k \le N-1)$
related to chained vectors.
\par
First, we show that the vectors $\mathbf{Q}_{ij}^{(l)}$
$(1 \le i < i+2 \le j \le N, \ l=n,n+1)$ can be expressed as the
functions of $\mathbf{Q}_{k,k+1}^{(l)}$ $(1 \le k \le N-1, \ l=n,n+1)$
related to the chain positional vectors $\mathbf{q}_{k,k+1}^{(l)}$ $(1
\le k \le N-1, \ l=n,n+1)$.
Because $\mathbf{Q}^{(l)} \in \mathcal{Q}$ $(l=n,n+1)$, where
$\mathcal{Q}$ is a $2(N-1)$-dimensional manifold defined by equation (21),
we can assume that the $(N-1)$ vectors $\mathbf{Q}_{k,k+1}^{(l)}$
$(1 \le k \le N-1,\ l = n, n+1)$,
which correspond to $2(N-1)$ variables, constitute a basis for the
manifold $\mathcal{Q}$.
In this basis, each vector $\mathbf{Q}_{ij}^{(l)} \in
\mathcal{Q}$ $(1 \le i < i+2 \le j \le N, \ l=n,n+1)$ related to a
non-chained position vector $\mathbf{q}_{ij}^{(l)}$ $(l=n,n+1)$
satisfies
\begin{eqnarray}
\mathbf{Q}_{ij}^{(n+1)} \mathbf{L} (\mathbf{Q}_{ij}^{(n+1)})^{\top}
= \sum_{k=i}^{j-1} \mathbf{Q}_{k,k+1}^{(n+1)} \mathbf{L}
(\mathbf{Q}_{k,k+1}^{(n+1)})^{\top}, \quad 1 \le i < i+2 \le j \le N.
\label{Proof_Q_{ij}^{(n+1)}}
\end{eqnarray}
In addition, to avoid cancellation of significant digits, the solutions of
equation (\ref{Proof_Q_{ij}^{(n+1)}}) for $\mathbf{Q}_{ij}^{(l)}$ $(l=n,n+1)$ are
written as follows:
{\footnotesize
\begin{eqnarray}
&& \hspace*{-15mm}
\mathbf{Q}_{ij}^{(l)} = \widetilde{\mathbf{Q}}_{ij}^{(l)} \nonumber\\
&& \hspace*{-15mm}
\equiv
\left\{\!\!
\begin{array}{cl}
\left(
\frac{\displaystyle \sqrt{2} \sum_{k=i}^{j-1} Q_{\!k,k+1[1]\!}^{(l)}
Q_{\!k,k+1[2]\!}^{(l)}}
{\displaystyle \sqrt{a_{ij}^{(l)} - b_{ij}^{(l)}}}, \
\displaystyle \sqrt{\frac{a_{ij}^{(l)} \!-\! b_{ij}^{(l)}}{2}}
\right)
& \mbox{if} \ b_{ij}^{(l)} < 0, \\
\displaystyle
\left(
\displaystyle \sqrt{\frac{a_{ij}^{(l)} \!+\! b_{ij}^{(l)}}{2}}, \
\frac{\displaystyle \sqrt{2} \sum_{k=i}^{j-1} Q_{\!k,k+1[1]\!}^{(l)}
Q_{\!k,k+1[2]\!}^{(l)}}
{\displaystyle \sqrt{a_{ij}^{(l)} + b_{ij}^{(l)}}}
\right)
& \mbox{if} \ b_{ij}^{(l)} \ge 0, \ 1 \le i < i
\!+\! 2 \le j \le N, \ l=n,n+1,
\end{array}
\right.
\label{tilde_Qij^{(n+1)}}
\end{eqnarray}
}
where
\begin{eqnarray*}
&& \displaystyle
a_{ij}^{(l)} \equiv \sqrt{
\left( \sum_{k=i}^{j-1} \left( (Q_{k,k+1[1]}^{(l)})^2 -
			 (Q_{k,k+1[2]}^{(l)})^2 \right) \right)^2 + 4
\left( \sum_{k=i}^{j-1} Q_{k,k+1[1]}^{(l)} Q_{k,k+1[2]}^{(l)}
\right)^2
}, \nonumber\\
&& \displaystyle
b_{ij}^{(l)} \equiv
\sum_{k=i}^{j-1} \left( (Q_{k,k+1[1]}^{(l)})^2 -
		  (Q_{k,k+1[2]}^{(l)})^2 \right),
\quad 1 \le i < i \!+\! 2 \le j \le N, \ l=n,n+1.
\end{eqnarray*}
\par
Next, we write the vectors $\mathbf{P}_{ij}^{(n+1/2)}$
$(1 \le i < i + 2 \le j \le N)$ related to the
non-chained momentum vectors $\mathbf{p}_{ij}^{(n+1/2)}$
as functions of
$\mathbf{P}_{k,k+1}^{(n+1/2)}$, $\mathbf{Q}_{k,k+1}^{(n)}$, and
$\mathbf{Q}_{k,k+1}^{(n+1)}$ $(1 \le k \le N-1)$.
$\mathbf{Q}_{ij}^{(n+1)} \in \mathcal{Q}$ $(1 \le i < i+2 \le j \le N)$
are expressed by equation (\ref{tilde_Qij^{(n+1)}}) using the basis constituted
by the $(N-1)$ vectors $\mathbf{Q}_{k,k+1}^{(n+1)}$ $(1 \le k \le N-1)$.
As described in Section 3.1, $\mathbf{Q}^{(n)} \in \mathcal{Q}$ is also
satisfied; namely, $\mathbf{Q}^{(n)}$ is restricted to the constraints
$\mathbf{\Phi} (\mathbf{Q}^{(n)}) = \mathbf{0}_{1 \times (N-1)(N-2)}$.
The form of these constraints is the same as that of the
constraints $\mathbf{\Phi} (\mathbf{Q}^{(n+1)}) = \mathbf{0}_{1 \times
(N-1)(N-2)}$, so $\mathbf{Q}^{(n)}$ satisfies the same form in
(\ref{Proof_Q_{ij}^{(n+1)}}):
\begin{eqnarray}
\mathbf{Q}_{ij}^{(n)} \mathbf{L} (\mathbf{Q}_{ij}^{(n)})^{\top}
= \sum_{k=i}^{j-1} \mathbf{Q}_{k,k+1}^{(n)} \mathbf{L}
(\mathbf{Q}_{k,k+1}^{(n)})^{\top}, \quad 1 \le i < i+2 \le j \le N.
\label{Proof_Q_{ij}^{(n)}}
\end{eqnarray}
Subtraction of equation (\ref{Proof_Q_{ij}^{(n)}}) from equation (\ref{Proof_Q_{ij}^{(n+1)}})
and subsequent division by
$2 \Delta t = 2 ( t^{(n+1)} - t^{(n)} )$ yield
\begin{eqnarray*}
&& \hspace*{-5mm} \displaystyle
\frac{1}{2 \Delta t} \!\left(\!
\mathbf{Q}_{ij}^{(n+1)} \mathbf{L} (\mathbf{Q}_{ij}^{(n+1)})^{\top}
\!-\! \mathbf{Q}_{ij}^{(n)} \mathbf{L} (\mathbf{Q}_{ij}^{(n)})^{\top}
\!\right)\!
\!=\! \sum_{k=i}^{j-1}
\frac{1}{2 \Delta t} \!\left(\!
\mathbf{Q}_{k,k+1}^{(n+1)} \mathbf{L}
(\mathbf{Q}_{k,k+1}^{(n+1)})^{\top}
\!-\! \mathbf{Q}_{k,k+1}^{(n)} \mathbf{L}
(\mathbf{Q}_{k,k+1}^{(n)})^{\top} \!\right)\!, \nonumber\\
&& \displaystyle \hspace*{110mm} 1 \le i < i+2 \le j \le N.
\end{eqnarray*}
This leads to
\begin{eqnarray}
\frac{\mathbf{Q}_{ij}^{(n+1)} - \mathbf{Q}_{ij}^{(n)}}{\Delta t}
\mathbf{L} (\mathbf{Q}_{ij}^{(n+1/2)})^{\top}
= \sum_{k=i}^{j-1}
\frac{\mathbf{Q}_{k,k+1}^{(n+1)} - \mathbf{Q}_{k,k+1}^{(n)}}{\Delta t}
\mathbf{L} (\mathbf{Q}_{k,k+1}^{(n+1/2)})^{\top},
\ 1 \le i < i+2 \le j \le N.
\label{Proof_New_Difference_Q_{ij}}
\end{eqnarray}
By using equations (\ref{dQ-redundant-GNBP}) and (\ref{tilde_Qij^{(n+1)}}),
equation (\ref{Proof_New_Difference_Q_{ij}}) is rewritten as
{\footnotesize
\begin{eqnarray}
&& \hspace*{-5mm} \displaystyle
\frac{m}{8 m_i m_j}
\frac{|\widetilde{\mathbf{Q}}_{ij}^{(n \!+\! 1)}|^2 \!+\! |\widetilde{\mathbf{Q}}_{ij}^{(n)}|^2}
{|\widetilde{\mathbf{Q}}_{ij}^{(n+1)}|^2 |\widetilde{\mathbf{Q}}_{ij}^{(n)}|^2}
\mathbf{P}_{ij}^{(n \!+\! 1/2)}
\mathbf{L} (\widetilde{\mathbf{Q}}_{ij}^{(n \!+\! 1/2)})^{\top}
\!=\! \sum_{k=i}^{j-1}
\frac{m}{8 m_k m_{k+1}}
\frac{|\mathbf{Q}_{k,k \!+\! 1}^{(n+1)}|^2 \!+\! |\mathbf{Q}_{k,k \!+\! 1}^{(n)}|^2}
{|\mathbf{Q}_{k,k \!+\! 1}^{(n+1)}|^2 |\mathbf{Q}_{k,k \!+\! 1}^{(n)}|^2}
\mathbf{P}_{k,k+1}^{(n \!+\! 1/2)}
\mathbf{L} (\mathbf{Q}_{k,k+1}^{(n \!+\! 1/2)})^{\top}, \nonumber\\
&& \hspace*{110mm} 1 \le i < i+2 \le j \le N.
\label{Proof_Relation_PQ}
\end{eqnarray}
}
We solve equation (\ref{Proof_Relation_PQ}) for $\mathbf{P}_{ij}^{(n+1/2)}$ $(1
\le i < i + 2 \le j \le N)$ and subsequently
substitute equation (\ref{tilde_Qij^{(n+1)}}) into the resulting solution.
Then, we give
\begin{eqnarray}
\mathbf{P}_{ij}^{(n+1/2)} =
\widetilde{\mathbf{P}}_{ij}^{(n+1/2)} \equiv
\left( \widetilde{P}_{ij[1]}^{(n+1/2)}, \widetilde{P}_{ij[2]}^{(n+1/2)}
\right), \ 1 \le i < i \!+\! 2 \le j \le N, \label{tilde_Pij^{(n+1/2)}}
\end{eqnarray}
where
{\footnotesize
\begin{eqnarray*}
\displaystyle \widetilde{P}_{ij[1]}^{(n+1/2)}
\!\!\!&=&\!\!\! m_i m_j
\frac{|\widetilde{\mathbf{Q}}_{ij}^{(n+1)}|^2
|\widetilde{\mathbf{Q}}_{ij}^{(n)}|^2}
{|\widetilde{\mathbf{Q}}_{ij}^{(n+1/2)}|^2
\left(\! |\widetilde{\mathbf{Q}}_{ij}^{(n+1)}|^2
+ |\widetilde{\mathbf{Q}}_{ij}^{(n)}|^2 \!\right)} \nonumber\\
&& \hspace*{-3mm} \displaystyle
\left(\!
\widetilde{Q}_{ij[1]}^{(n+1/2)}
\sum_{k=i}^{j-1}
\frac{\left(\! |\mathbf{Q}_{k,k+1}^{(n+1)}|^2 \!+\!
|\mathbf{Q}_{k,k+1}^{(n)}|^2 \!\right)
\left(\!
P_{k,k+1[1]}^{(n+1/2)} Q_{k,k+1[1]}^{(n+1/2)}
\!-\! P_{k,k+1[2]}^{(n+1/2)} Q_{k,k+1[2]}^{(n+1/2)} \!\right)}
{m_k m_{k+1} |\mathbf{Q}_{k,k+1}^{(n+1)}|^2
|\mathbf{Q}_{k,k+1}^{(n)}|^2}
\right. \nonumber\\
&& \hspace*{-3mm}
\left. \quad
+ \widetilde{Q}_{ij[2]}^{(n+1/2)}
\sum_{k=i}^{j-1} \frac{\left(\! |\mathbf{Q}_{k,k+1}^{(n+1)}|^2 \!+\!
			|\mathbf{Q}_{k,k+1}^{(n)}|^2 \!\right)
\left(\!
P_{k,k+1[1]}^{(n+1/2)} Q_{k,k+1[2]}^{(n+1/2)} \!+\! P_{k,k+1[2]}^{(n+1/2)}
Q_{k,k+1[1]}^{(n+1/2)} \!\right)}
{m_k m_{k+1} |\mathbf{Q}_{k,k+1}^{(n+1)}|^2 |\mathbf{Q}_{k,k+1}^{(n)}|^2}
\!\right), \nonumber\\
\displaystyle \widetilde{P}_{ij[2]}^{(n+1/2)}
\!\!\!&=&\!\!\! m_i m_j
\frac{|\tilde{\mathbf{Q}}_{ij}^{(n+1)}|^2
|\tilde{\mathbf{Q}}_{ij}^{(n)}|^2}
{|\tilde{\mathbf{Q}}_{ij}^{(n+1/2)}|^2
\left( |\tilde{\mathbf{Q}}_{ij}^{(n+1)}|^2
+ |\tilde{\mathbf{Q}}_{ij}^{(n)}|^2\right)} \nonumber\\
&& \hspace*{-3mm} \left(\!
\tilde{Q}_{ij[1]}^{(n+1/2)}
\sum_{k=i}^{j-1} \frac{\left(\! |\mathbf{Q}_{k,k+1}^{(n+1)}|^2 \!+\!
			|\mathbf{Q}_{k,k+1}^{(n)}|^2 \!\right) \left(\!
P_{k,k+1[1]}^{(n+1/2)} Q_{k,k+1[2]}^{(n+1/2)} \!+\! P_{k,k+1[2]}^{(n+1/2)}
Q_{k,k+1[1]}^{(n+1/2)} \!\right)}{m_k m_{k+1}
|\mathbf{Q}_{k,k+1}^{(n+1)}|^2 |\mathbf{Q}_{k,k+1}^{(n)}|^2} \right. \nonumber\\
&& \hspace*{-3mm}
\left.
\quad - \tilde{Q}_{ij[2]}^{(n+1/2)}
\sum_{k=i}^{j-1} \frac{\left(\! |\mathbf{Q}_{k,k+1}^{(n+1)}|^2 \!+\!
			|\mathbf{Q}_{k,k+1}^{(n)}|^2 \!\right)
\left(\!
P_{k,k+1[1]}^{(n+1/2)} Q_{k,k+1[1]}^{(n+1/2)} \!-\! P_{k,k+1[2]}^{(n+1/2)}
Q_{k,k+1[2]}^{(n+1/2)} \!\right)}
{m_k m_{k+1} |\mathbf{Q}_{k,k+1}^{(n+1)}|^2 |\mathbf{Q}_{k,k+1}^{(n)}|^2}
\!\right), \nonumber\\
&& \hspace*{90mm}
1 \le i < i \!+\! 2 \le j \le N.
\end{eqnarray*}
}
Because of the definition of the midpoint value $(\bullet)^{(n+1/2)}$ and
equation (\ref{tilde_Pij^{(n+1/2)}}),
\begin{eqnarray}
\mathbf{P}_{ij}^{(n+1)} = \widetilde{\mathbf{P}}_{ij}^{(n+1)}
= 2 \widetilde{\mathbf{P}}_{ij}^{(n+1/2)} - \widetilde{\mathbf{P}}_{ij}^{(n)}\
(1 \le i < j \le N). \label{Pij^{(n+1)}}
\end{eqnarray}
\par
Using equations (\ref{tilde_Qij^{(n+1)}}) and (\ref{tilde_Pij^{(n+1/2)}}), we can
rewrite equation (\ref{dP-redundant-GNBP}) in the d-RG$N$BP as
{\footnotesize
\begin{eqnarray}
\displaystyle
\sum_{i=1}^{j-2} \tilde{\mathbf{G}}_{ij}^{(n+1)} - \sum_{i=j+2}^N
\tilde{\mathbf{G}}_{ji}^{(n+1)}
= - \mathbf{G}_{j-1,j}^{(n+1)} + \mathbf{G}_{j,j+1}^{(n+1)}, \
j=2,\cdots,N-1; \quad\sum_{i=1}^{N-2} \tilde{\mathbf{G}}_{i,N}^{(n+1)} =
- \mathbf{G}_{N-1,N}^{(n+1)},
\label{Modified_Relation-tilde_G}
\end{eqnarray}
}
where
{\footnotesize
\begin{eqnarray}
&& \hspace*{-5mm} \tilde{\mathbf{G}}_{ij}^{\!(n \!+\! 1)\!} \nonumber\\
&& \hspace*{-5mm} \equiv \displaystyle
\frac{1}
{2 |\tilde{\mathbf{Q}}_{ij}^{\!(n \!+\! 1/2)\!}|^2}
\!\left(\!
\frac{\tilde{\mathbf{P}}_{ij}^{\!(n \!+\! 1)\!} \!-\! \tilde{\mathbf{P}}_{ij}^{(n)}}
{\Delta t} \!-\! \frac{1}{|\tilde{\mathbf{Q}}_{ij}^{\!(n \!+\! 1)\!}|^2
|\tilde{\mathbf{Q}}_{ij}^{(n)}|^2}
\!\left(\!
\frac{m}{8 m_i m_j}
\!\left(\! |\tilde{\mathbf{P}}_{ij}^{\!(n \!+\! 1)\!}|^2 \!+\!
   |\tilde{\mathbf{P}}_{ij}^{(n)}|^2 \!\right)\!
\!-\! 2 m_i m_j
\!\right)\! \tilde{\mathbf{Q}}_{ij}^{(n \!+\! 1/2)}
\!\right)\! \mathbf{L} \left(\! \tilde{\mathbf{Q}}_{ij}^{\!(n \!+\! 1/2)\!}
		       \!\right)^{\top}, \nonumber\\
&& \hspace*{110mm} 1 \le i < i + 2 \le j \le N,
\label{def-dGij-nonchained}
\end{eqnarray}
}
and $\mathbf{G}_{j-1,j}^{(n+1)}$, $\mathbf{G}_{j,j+1}^{(n+1)}$, and
$\mathbf{G}_{N-1,N}^{(n+1)}$ have the same form as equation (\ref{def-dGij}).
Note that equation (\ref{def-dGij-nonchained}) is described by only $4(N-1)$
vectors, $\mathbf{Q}_{k,k+1}^{(n)}$, $\mathbf{Q}_{k,k+1}^{(n+1)}$,
$\mathbf{P}_{k,k+1}^{(n)}$, and $\mathbf{P}_{k,k+1}^{(n+1)}$.
In addition, these vectors satisfy equation (\ref{dQ-redundant-GNBP}); namely,
\begin{eqnarray}
\displaystyle
\frac{\mathbf{Q}_{k,k+1}^{(n+1)} - \mathbf{Q}_{k,k+1}^{(n)}}{\Delta t}
= \frac{m}{8 m_k m_{k+1}}
\frac{|\mathbf{Q}_{k,k+1}^{(n+1)}|^2 +
|\mathbf{Q}_{k,k+1}^{(n)}|^2}{|\mathbf{Q}_{k,k+1}^{(n+1)}|^2
|\mathbf{Q}_{k,k+1}^{(n)}|^2} \mathbf{P}_{k,k+1}^{(n+1/2)}, \ 1 \le k \le N-1.
\label{def-modified-dFi,i+1}
\end{eqnarray}
We call the discrete-time system composed of equations (\ref{Modified_Relation-tilde_G})
and (\ref{def-modified-dFi,i+1}) the {\itshape discrete-time chain regularization of the G$N$BP (d-CRG$N$BP)}.
The d-CRG$N$BP describes the same motion as the d-RG$N$BP
(35), as shown by the following lemma.
\begin{lmm}
Suppose
\begin{enumerate}
\item[(i)] The $(N-1)(N-2)$ vectors $\mathbf{Q}_{ij}^{(n+1)}$ and
$\mathbf{P}_{ij}^{(n+1)}$ $(1 \le i < i+2 \le j \le N)$ are given by
	    equations (\ref{tilde_Qij^{(n+1)}}) and (\ref{Pij^{(n+1)}}).
\item[(ii)] The $2(N-1)$ vectors $\mathbf{Q}_{k,k+1}^{(n+1)}$ and
$\mathbf{P}_{k,k+1}^{(n+1)}$ $(1 \le k \le N-1)$ are the solutions of
the d-CRG$N$BP composed of equations (\ref{Modified_Relation-tilde_G})
and (\ref{def-modified-dFi,i+1}).
\end{enumerate}
Then, the $N(N-1)$ vectors $\mathbf{Q}_{ij}^{(n+1)}$ and $\mathbf{P}_{ij}^{(n+1)}$
$(1 \le i < j \le N)$ satisfy the d-RG$N$BP (35).
\end{lmm}
{\itshape Proof.} \ \\
(a)\ {\itshape Derivation of equation (\ref{dQ-redundant-GNBP})}
\ \\
The vectors $\mathbf{P}_{ij}^{(n+1/2)}$ $(1 \le i < i+2 \le j \le N)$
in equation (\ref{tilde_Pij^{(n+1/2)}}) satisfy equation (\ref{Proof_Relation_PQ}).
Further, through equation (\ref{def-modified-dFi,i+1}), equation (\ref{Proof_Relation_PQ})
leads to
\begin{eqnarray}
&& \hspace*{-5mm} \displaystyle
\frac{m}{8 m_i m_j}
\frac{|\mathbf{Q}_{ij}^{(n \!+\! 1)}|^2 \!+\! |\mathbf{Q}_{ij}^{(n)}|^2}
{|\mathbf{Q}_{ij}^{(n+1)}|^2 |\mathbf{Q}_{ij}^{(n)}|^2}
\mathbf{P}_{ij}^{(n \!+\! 1/2)}
\mathbf{L} (\mathbf{Q}_{ij}^{(n \!+\! 1/2)})^{\top}
\!=\! \sum_{k=i}^{j-1}
\frac{\mathbf{Q}_{k,k+1}^{(n+1)} - \mathbf{Q}_{k,k+1}^{(n)}}{\Delta t}
\mathbf{L} (\mathbf{Q}_{k,k+1}^{(n \!+\! 1/2)})^{\top}, \nonumber\\
&& \hspace*{100mm} 1 \le i < i+2 \le j \le N.
\label{modified_Proof_Relation_PQ}
\end{eqnarray}
Similarly, the vectors $\mathbf{Q}_{ij}^{(n+1)}$ $(1 \le i < i+2 \le j
\le N)$ defined by equation (\ref{tilde_Qij^{(n+1)}}) fulfill
equation (\ref{Proof_Q_{ij}^{(n+1)}}) and $\mathbf{Q}_{ij}^{(n+1)} \in
\mathcal{Q}$, so equation (\ref{Proof_New_Difference_Q_{ij}}) is also fulfilled.
Because the r.h.s. of equation (\ref{modified_Proof_Relation_PQ}) coincides with
that of equation (\ref{Proof_New_Difference_Q_{ij}}), equation (\ref{dQ-redundant-GNBP})
in the d-RG$N$BP is given.
\ \\
\ \\
(b)\ {\itshape Derivation of equation (\ref{dP-redundant-GNBP})}
\ \\
Substitution of equations (\ref{tilde_Qij^{(n+1)}}),
(\ref{tilde_Pij^{(n+1/2)}}), and (\ref{Pij^{(n+1)}}) into
equation (\ref{dP-redundant-GNBP}) yields equation (\ref{Modified_Relation-tilde_G}).
Namely, equation (\ref{Modified_Relation-tilde_G}), which is part of
condition (ii), equals equation (\ref{dP-redundant-GNBP}) under condition
(i).
Accordingly, $\mathbf{Q}_{ij}^{(n+1)}$ and $\mathbf{P}_{ij}^{(n+1)}$
$(1 \le i < j \le N)$ satisfying conditions (i) and (ii) follow
from equation (\ref{dP-redundant-GNBP}).
\ \\
\ \\
(c)\ {\itshape Derivation of equation (\ref{dPhi-redundant-GNBP})}
\ \\
We have already stated that equation (\ref{tilde_Qij^{(n+1)}}) is the solution of
equation (\ref{Proof_Q_{ij}^{(n+1)}}).
Substitution of equation (\ref{Proof_Q_{ij}^{(n+1)}}) into the l.h.s. of
equation (\ref{dPhi-redundant-GNBP}) yields $\mathbf{0}$.
Thus, $\mathbf{Q}_{ij}^{(n+1)}$ $(1 \le i < j \le N)$ satisfy
equation (\ref{dPhi-redundant-GNBP}) under conditions (i) and (ii).
\quad \fbox{}
\ \\
\par
Lemma 2 clarifies that the d-CRG$N$BP reproduces the motion of the
d-RG$N$BP, and Theorem 1 shows that the d-RG$N$BP preserves all the conserved
quantities but the angular momentum.
Therefore, the conservation of quantities in the d-CRG$N$BP is stated in the following
theorem.
\begin{thm}
(Conserved quantities of d-CRG$N$BP) \
The d-CRG$N$BP composed of equations (\ref{Modified_Relation-tilde_G}) and
(\ref{def-modified-dFi,i+1}) exactly preserves the following three
conserved quantities:
\begin{enumerate}
\item[1.] the Hamiltonian defined by equation (\ref{Sec2.1:CQ.H}),
\item[2.] the linear momentum $\displaystyle \mathbf{l} \equiv \sum_{i=1}^N
	  \mathbf{p}_i' = \mathbf{0}$,
\item[3.] the position of the center of mass $\displaystyle \mathbf{c}
	  \equiv \sum_{i=1}^N m_i \mathbf{q}_i' = \mathbf{0}$.
\end{enumerate}
\end{thm}
\section{Numerical Results}
In this section, we compare the results obtained with the following methods:
\begin{enumerate}
\item[]{\itshape RK$4$:} \
The fourth-order Runge--Kutta method, which is used for integrating
equation (\ref{Sec2.1:CQ.diff2-q}),
\item[]{\itshape SI$4$:} \
The fourth-order symplectic method, which is applied to
equation (\ref{Sec2.1:CQ.diff2-q}),
\item[]{\itshape G:} \
Greenspan's energy conserving method \citep{Greenspan,LaBudde}, which is
second-order accurate and is applied to equation (\ref{Sec2.1:CQ.diff2-q}),
\item[]{\itshape d-CRG$N$BP:} \
The d-CRG$N$BP, which is given by equations (\ref{Modified_Relation-tilde_G})
and (\ref{def-modified-dFi,i+1}), and is second-order accurate.
\end{enumerate}
\par
In Section 4.1, we show that the d-CRG$N$BP precisely conserves the
Hamiltonian of the general four-body problem (G$4$BP) for a long
integration interval.
Next, in Section 4.2, we show that the d-CRG$N$BP computes
equilibrium solutions and periodic orbits around equilibrium points in
the G$3$BP, G$4$BP, and general five-body problem (G$5$BP) more
correctly than the other methods.
\subsection{Conservation}
First, let us show that the d-CRG$N$BP preserves the Hamiltonian
$H$ exactly and the angular momentum $\mathbf{j} = \sum_{i=1}^4
\mathbf{p}_{i} \times \mathbf{q}_{i}$ approximately.
Figure 1 shows the dependence of the relative error growth of the
Hamiltonian $H$ and angular momentum $\mathbf{j}$ for the G$4$BP on the
RK$4$, SI$4$, G and d-CRG$N$BP methods, respectively.
The adopted initial conditions are as follows:
\begin{eqnarray}
&& \hspace*{-12mm}
m_1 = m_2 = m_3 = m_4 = 0.25, \nonumber\\
&& \hspace*{-12mm}
\displaystyle \mathbf{p}'_1 = (0, 0.1), \
\mathbf{p}'_2 = (0, -0.05), \
\mathbf{p}'_3 = (0, -0.1), \
\mathbf{p}'_4 = (0, 0.05), \nonumber\\
&& \hspace*{-12mm}
\mathbf{q}'_1 = (-10, 0), \
\mathbf{q}'_2 = (-12, 0), \
\mathbf{q}'_3 = (10, 0), \
\mathbf{q}'_4 = (12, 0).
\label{Quasi-FA}
\end{eqnarray}
In addition, the step size is fixed at $\Delta t=0.1$.
The initial condition corresponds to the Caledonian symmetric four-body
problem, in which the four bodies are always configured in a
parallelogram \citep{Szell}.
Each method always sets the linear momentum $\mathbf{l}$ and the center
of mass $\mathbf{c}$ at the origin in the barycentric frame.
\par
For the RK$4$ method, the relative error of $H_{\mbox{\footnotesize rel}}$
grows with time $t$, whereas the error is bounded by a sufficiently small
value $10^{-7}$ for the SI$4$, G, and d-CRG$N$BP methods.
In particular, the G and d-CRG$N$BP methods conserve $H$ with $10^{-15}$
accuracy.
In addition, the error of $\mathbf{j}$ grows in proportion to the time $t$
for the RK$4$ method, whereas it is bounded by $10^{-5}$ for the SI$4$, G, and
d-CRG$N$BP methods.
Specifically, the SI$4$ and G methods precisely keep $\mathbf{j}$.
Because only Greenspan's energy conserving method (G) preserves both
$H$ and $\mathbf{j}$, it would appear that this method reproduces orbits
more precisely than the others.
However, Section 4.2 will clarify that this prediction is incorrect.
\subsection{Periodic Orbits}
In Section 4.1, we showed that the d-CRG$N$BP does not exactly
preserve the angular momentum.
Therefore, it is questionable whether the d-CRG$N$BP can reproduce
various orbits of the G$3$BP and G$4$BP because the orbits lie on the
manifold determined by conserved quantities.
To answer this question, we show that various orbits computed by the
d-CRG$N$BP accurately coincide with those of the G$3$BP, G$4$BP and
G$5$BP.
\subsubsection{Choreographies in the G$3$BP}
In choreography solutions, all the bodies are equally spaced along a
single closed orbit.
The three-body figure-eight choreography was discovered by Chenciner and
Montgomery \citep{Chenciner} and located numerically by Sim{\'o}
\citep{Simo-2000}.
The initial conditions are those cited in \citep{Simo-2000}:
\begin{eqnarray}
&&
m_1 = m_2 = m_3 = 1, \nonumber\\
&& \displaystyle
\mathbf{p}'_1 = (0.46620369, \ 0.43236573), \
\mathbf{p}'_2 = (-0.93240737, \ -0.86473146), \nonumber\\
&& \displaystyle
\mathbf{p}'_3 = (0.46620369, \ 0.43236573), \
\mathbf{q}'_1 = (0.97000436, \ -0.24308753), \nonumber\\
&& \displaystyle
\mathbf{q}'_2 = (0, \ 0), \
\mathbf{q}'_3 = (-0.97000436, \ 0.24308753).
\label{choreography-G3BP}
\end{eqnarray}
Applying the RK$4$, SI$4$, G, and d-CRG$N$BP methods, we obtained the
orbits of particle $m_3$ in the barycentric frame.
We used the common time step $\Delta t = 0.1$ and integrated until
$t_f = 10,000$ (see Figure 2).
Particle $m_3$ theoretically follows a closed figure-eight orbit under
this condition and travels more than $1500$ times around the orbit.
The SI$4$ and d-CRG$N$BP methods give the closed figure-eight orbit with
high precision, whereas the RK$4$ and G methods do not obtain a closed
orbit.
In particular, the RK$4$ method shows the orbit of particle $m_3$ shifting
away from the figure-eight orbit, and the G method obtains an orbit drifting
around that orbit.
\subsubsection{Stable Equilibrium Points in the G$4$BP}
We clarify that the d-CRG$N$BP precisely computes some stable equilibrium
solutions in the circular G$4$BP which has four finite masses $m_1$,
$m_2$, $m_3$ and $m_4$. 
As the mass $m_4$ goes to zero, the equilibrium solutions in the
circular G$4$BP reduce to those in the circular restricted four-body
problem (CR$4$BP), in each of which the three masses $m_1$, $m_2$ and
$m_3$ form an equilateral triangle and orbit a common circle according
to the Lagrangian solution \citep{BP, Majorana} in the barycentric
inertial frame.
The CR$4$BP has eight equilibrium points, at one of which the massless
particle $m_4$ rests in a rotating frame where the three primaries, $m_1$,
$m_2$, and $m_3$ are fixed at the vertices of an equilateral triangle.
Two of the equilibrium points are linearly stable \citep{BP,
Majorana}.
\par
The G$4$BP also has two stable quasi-equilibrium solutions corresponding
to these two equilibrium points.
The initial conditions are given for one of the two quasi-equilibrium
solutions as
\begin{eqnarray}
&& m_1 = 0.01, \ m_2 = 0.021, \ m_3 = 0.969, \ m4 = 1 \times 10^{-12}, 
\nonumber\\
&& \mathbf{p}'_1 = \left( -6.3263361598479256638 \times 10^{-15}, \
		     0.97966882159151619201 \times 10^{-2} \right), \nonumber\\
&& \mathbf{p}'_2 = \left( -0.17988477894964014249 \times 10^{-1}, \ 
		     0.97372405753200600321 \times 10^{-2} \right), \nonumber\\
&& \mathbf{p}'_3 = \left( 0.17988477894362678026 \times 10^{-1}, \
		     -0.19533928791999029945 \times 10^{-1} \right), \nonumber\\
&& \mathbf{p}'_4 = \left( 6.3263361598479256638 \times 10^{-13}, \ 
		     7.6380799288046759201\times 10^{-13} \right), 
\nonumber\\
&& \mathbf{q}'_1 = \left( 0.97966882159151279741, \ 
		    6.3263871598479246128 \times 10^{-13} \right), \nonumber\\
&& \mathbf{q}'_2 = \left( 0.46367812263428759741, \ 
		    0.85659418547566503871 \right), \nonumber\\
&& \mathbf{q}'_3 = \left( -0.20158853242517002593\times 10^{-1}, \
		     -0.18563960675296861284 \times 10^{-1} \right), \nonumber\\
&& \mathbf{q}'_4 = \left( 0.76380799288046499741, \ 
		    -0.63263361598479246128 \right).
\label{equilibrium-CG4BP}
\end{eqnarray}
\par
We introduce a rotating frame $\mbox{O}-x'_{[1]} x'_{[2]}$.
In this frame, the origin stays at the center of mass, and the $x'_{[1]}$
axis passes through the origin and primary $m_1$.
Theoretically, the position of particle $m_4$, $\mathbf{x}'_4$, is
almost at rest in the frame as well as those of the primaries, $m_1$,
$m_2$, and $m_3$.
Since the G and d-CRG$N$BP methods are all implicit schemes, Newton's
method is necessary for solving them.
Here, we give starting values by Heun's method for Newton's method and
for a convergence tolerance (as measured by the norm of the difference
between successive iterates) of less than $10^{-16}$.
In the case of $\Delta t = 0.1$, Newton's iteration in the d-CRG$N$BP
method converges, while that in the G method does not converge.
For the RK$4$, SI$4$, and d-CRG$N$BP methods, Figure 3 shows $|\Delta
\mathbf{x}'_4(t)|$, which is the shift in $\mathbf{x}'_4 (t)$ at time $t$
from the initial position $\mathbf{x}'_4 (0)$.
Applying the RK$4$, SI$4$, and d-CRG$N$BP methods, we computed $|\Delta
\mathbf{x}'_4(t)|$ in the rotating frame $\mbox{O}-x'_{[1]}x'_{[2]}$.
We used the common time step $\Delta t = 0.1$ and integrated until $t_f
= 1,000,000$.
The result shown in Figure 3 indicates that the shift $|\Delta
\mathbf{x}'_4(t)|$ grows with time for all of the RK$4$ and d-CRG$N$BP
methods and that the lower limit of error of SI$4$ increases.
Also, the shift $|\Delta \mathbf{x}'_4(t)|$ for the d-CRG$N$BP is least
for the time interval $0 \le t \le t_f$.
Consequently, the result appears that the d-CRG$N$BP method reproduces
the equilibrium solution more precisely than the others.
%
%
%
\subsubsection{Stable Equilibrium Points in the G$5$BP}
Further, we clarify that the d-CRG$N$BP accurately computes two
stable stationary configurations in the $1 + 4$-body problem.
Such stable configurations are given as those for $n=4$ in the
$1 + n$-body problem with one large mass and $n$ small masses.
For an arbitrary integer $n = 2,3, \cdots$, the relations satisfied by the
stationary stable configurations in the $1 + n$-body problem
are described, and, for some integers, stationary configurations are
numerically obtained from these relations \citep{Casasayas, Cors, Salo}.
The initial condition corresponds to one of those cited (see
the last column of Table III in \cite{Cors}), as follows:
\begin{eqnarray}
&& m_1=1.0, \ m_2 = m_3 = m_4 = m_5 = 1.0 \times 10^{-8}, \nonumber\\
&& \mathbf{p}'_1 = \left( -0.0000000250666972, \ -0.0000000143699690
		   \right), \nonumber\\
&& \mathbf{p}'_2 = \left( 0.0000000086294301, \ -0.0000000050530123
		   \right), \nonumber\\
&& \mathbf{p}'_3 = \left( 0.0000000098113650, \ 0.0000000019331608
		   \right), \nonumber\\
&& \mathbf{p}'_4 = \left( 0.0000000066259023, \ 0.0000000074898207
		   \right), \nonumber\\
&& \mathbf{p}'_5 = \left( -0.0000000000000003, \ 0.0000000099999999
		   \right), \nonumber\\
&& \mathbf{q}'_1 = \left( -0.0000000143699690, \ 0.0000000250666972
		   \right), \nonumber\\
&& \mathbf{q}'_2 = \left( -0.5053012275872134, \ -0.8629430112463499
		   \right), \nonumber\\
&& \mathbf{q}'_3 = \left( 0.1933160775788700, \ -0.9811365039615660
		   \right), \nonumber\\
&& \mathbf{q}'_4 = \left( 0.7489820652276899, \ -0.6625902287414661
		   \right), \nonumber\\
&& \mathbf{q}'_5 = \left( 0.9999999856300310, \ 0.0000000250614611
		   \right).
\label{Central-Configuration-1+4-BP}
\end{eqnarray}
Because $m_i$ $(i=2,\cdots,5)$ is sufficiently small, the stationary
configuration corresponding to the initial condition
(\ref{Central-Configuration-1+4-BP}) is stable.
In the same rotating frame $\mbox{O}-x'_{[1]}x'_{[2]}$ in Section 4.2.2,
each position of mass $m_i$ $(i = 1, 2, \cdots, 5)$,
$\mathbf{x}_i (t)$ is theoretically fixed at an arbitrary time $t$.
We define $e_{\footnotesize \mbox{max}} (t)$ as the maximum of five
differences:
$\left| \mathbf{x}_1 (t) - \mathbf{x}_1 (0) \right|$,
$\left| \mathbf{x}_2 (t) - \mathbf{x}_2 (0) \right|$, $\cdots$,
$\left| \mathbf{x}_5 (t) - \mathbf{x}_5 (0) \right|$.
Theoretically, $e_{\footnotesize \mbox{max}} (t)$ is zero;
$e_{\footnotesize \mbox{max}} (t)$ is actually nonzero
because of perturbation by the small masses and the influence of the numerical
error.
We computed $e_{\footnotesize \mbox{max}} (t)$ using the RK$4$, SI$4$, G,
and d-CRG$N$BP methods (see Figure 4).
We used the common time step $\Delta t = 0.01$ and integrated until $t_f
= 10,000$.
The result indicates that the error $e_{\footnotesize \mbox{max}} (t)$
grows linearly with time for the RK$4$ method, whereas the SI$4$, G, and
d-CRG$N$BP methods keep $e_{\footnotesize \mbox{max}} (t)$ within
$4 \times 10^{-4}$.
Thus, only the RK$4$ method cannot retain the central configuration in the
$1+4$-body problem.
The reason is that the errors of both the Hamiltonian $H$ and angular
momentum $\mathbf{j}$ increase with time $t$ for the RK$4$.
\subsubsection{Periodic Orbits around Equilibrium Points in the G$4$BP}
Finally, we show that the d-CRG$N$BP accurately computes some
quasi-periodic orbits that reduce to periodic ones in the CR$4$BP
\citep{Baltagiannis-b} as the mass $m_4$ goes to zero.
Baltagiannis and Papadakis listed the initial states for the non-symmetric
periodic orbits of nine families in the CR$4$BP, all of which are
linearly stable \citep{Baltagiannis-b}.
If the mass $m_4$ is sufficiently low, the G$4$BP has similar families.
We give the initial state for one of the families in the G$4$BP as
follows:
\begin{eqnarray}
&& m_1=0.97, \ m_2=0.02, \ m_3=0.01, m_4=2 \times 10^{-17}, \nonumber\\
&& \mathbf{p}'_1 = \left(0, \ 0.0256637877173266
		   \right), \
\mathbf{p}'_2 = \left(-0.0065465367070798, \ -0.0183690733882485
		   \right), \nonumber\\
&& \mathbf{p}'_3 = \left(0.0065465367070798 , \ -0.0072947143290781
		   \right), \nonumber\\
&& \mathbf{p}'_4 = \left(3.292738 \times 10^{-19}, \
		 -9.55592100000001 \times 10^{-18}
		   \right), \nonumber\\
&& \mathbf{q}'_1 = \left(0.026457513110646, \ 0
		   \right), \
\mathbf{q}'_2 = \left(-0.9184536694124222, \  0.3273268353539883
		   \right), \nonumber\\
&& \mathbf{q}'_3 = \left(-0.7294714329078083, \ -0.6546536707079774
		   \right), \
\mathbf{q}'_4 = \left(-1.55780271, \ 0
		   \right).
\label{Periodic-EG4BP}
\end{eqnarray}
As $m_4 \to 0$, this initial state corresponds to that for family $f_7$
in the CR$4$BP, where the primaries $m_1$, $m_2$, and $m_3$ revolve in a
circle at the angular velocity $\omega=1$ (see Table 2 in
\cite{Baltagiannis-b}).
Under the extremely small effect of the gravity of mass
$m_4$, the three primaries move in approximately the same circle.
Therefore, we introduce a frame $\mbox{O}-{x'_{[1]} x'_{[2]}}$ rotating
around the center of mass with the angular velocity $\omega = 1$, where the
$x'_{[1]}$ axis always passes through the origin and the primary $m_1$.
In the frame $\mbox{O}-{x_{[1]} x_{[2]}}$, the three primaries move very little,
in principle.
In addition, Baltagiannis and Papadakis numerically clarified that
mass $m_4$ travels in a non-symmetric closed orbit for the initial state
(\ref{Periodic-EG4BP}) as $m_4 \to 0$.
\par
Applying the RK$4$, SI$4$, G, and d-CRG$N$BP methods, we computed the
orbit of particle $m_4$ in this frame (see Figure 5).
Using a fixed time step $\Delta t = 0.1$ for the RK$4$, SI$4$, G, and
d-CRG$N$BP methods, we integrated over the time interval $0 \le t \le
10,000$.
For the d-CRG$N$BP, particle $m_4$ moves along a perturbed orbit
around the closed non-symmetric orbit, whereas it escapes for the
RK$4$, SI$4$, and G methods.
In particular, Greenspan's energy-conserving method (G) does
not give the closed orbit, even though it is the only one to precisely
conserve the Hamiltonian $H$ and angular momentum $\mathbf{j}$.
\par
The results given in Sections 4.2.2, 4.2.3, and 4.2.4 show that (i) only
the d-CRG$N$BP can compute an equilibrium solution of an elliptic G$4$BP
and periodic orbits around equilibrium points with high accuracy, and that
(ii) the conservation of both $H$ and $\mathbf{j}$ is not necessarily
sufficient for obtaining the orbit of particle $m_4$.
\section{Conclusion}
We applied a chain regularization method and an extension of the
d'Alembert-type scheme \citep{Betsch2005} to the general
$N$-body problem.
Then, we presented a discrete-time chain regularization of the general $N$-body
problem (d-CRG$N$BP), which includes the discrete-time general three-body
problem (d-G$3$BP) proposed by the author \citep{Minesaki-2013a}.
The d-CRG$N$BP is second-order accurate and theoretically keeps all
the conserved quantities but the angular momentum.
For $N = 3$ and $4$, Figure 1 in this article and Figure
2 in \citep{Minesaki-2013a} show that the d-CRG$N$BP preserves the
Hamiltonian.
\par
For $N = 3$ and $4$, the numerical results demonstrate that the d-CRG$N$BP method
is superior to the symplectic and energy-momentum methods in the
following sense: only the d-CRG$N$BP can reproduce all the equilibrium points
and periodic orbits precisely, whereas the fourth-order Runge--Kutta,
symplectic, and second-order energy momentum methods cannot reproduce all of
them.
In particular, for $N = 3$, the author has already proved that
the d-CRG$N$BP method has the same equilibrium points as the original
general three-body problem \citep{Minesaki-2013b}.
Further, for $N = 5$, the d-CRG$N$BP method as well as the symplectic
and energy-momentum methods can precisely give all the stable equilibrium
points.
In future papers, we will analytically clarify that the d-CRG$N$BP has the
same equilibrium points in the restricted four-body problem and the
circular $N$-body problem.

\begin{figure}
\epsscale{.80}
\plotone{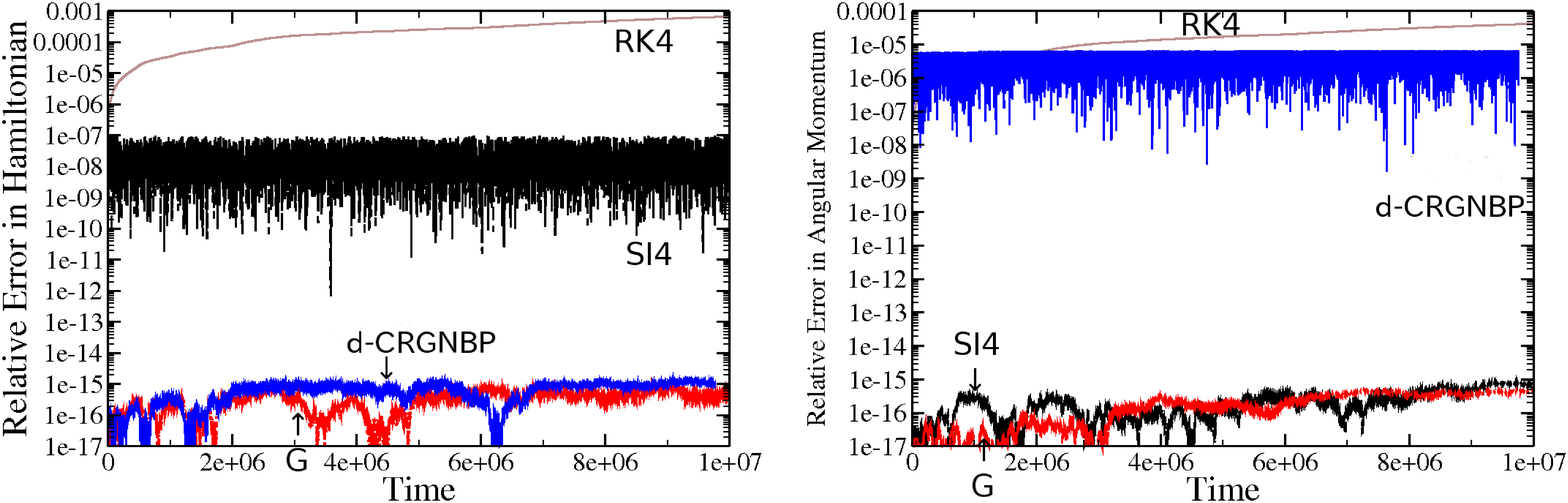}
\caption{
Relative errors of Hamiltonian $H$ and angular momentum
 $\mathbf{j}$ given by RK$4$, SI$4$, G, and d-CRG$N$BP methods.
A Caledonian symmetric four-body problem is integrated \citep{Szell}
}
\end{figure}
\begin{figure}
\epsscale{.80}
\plotone{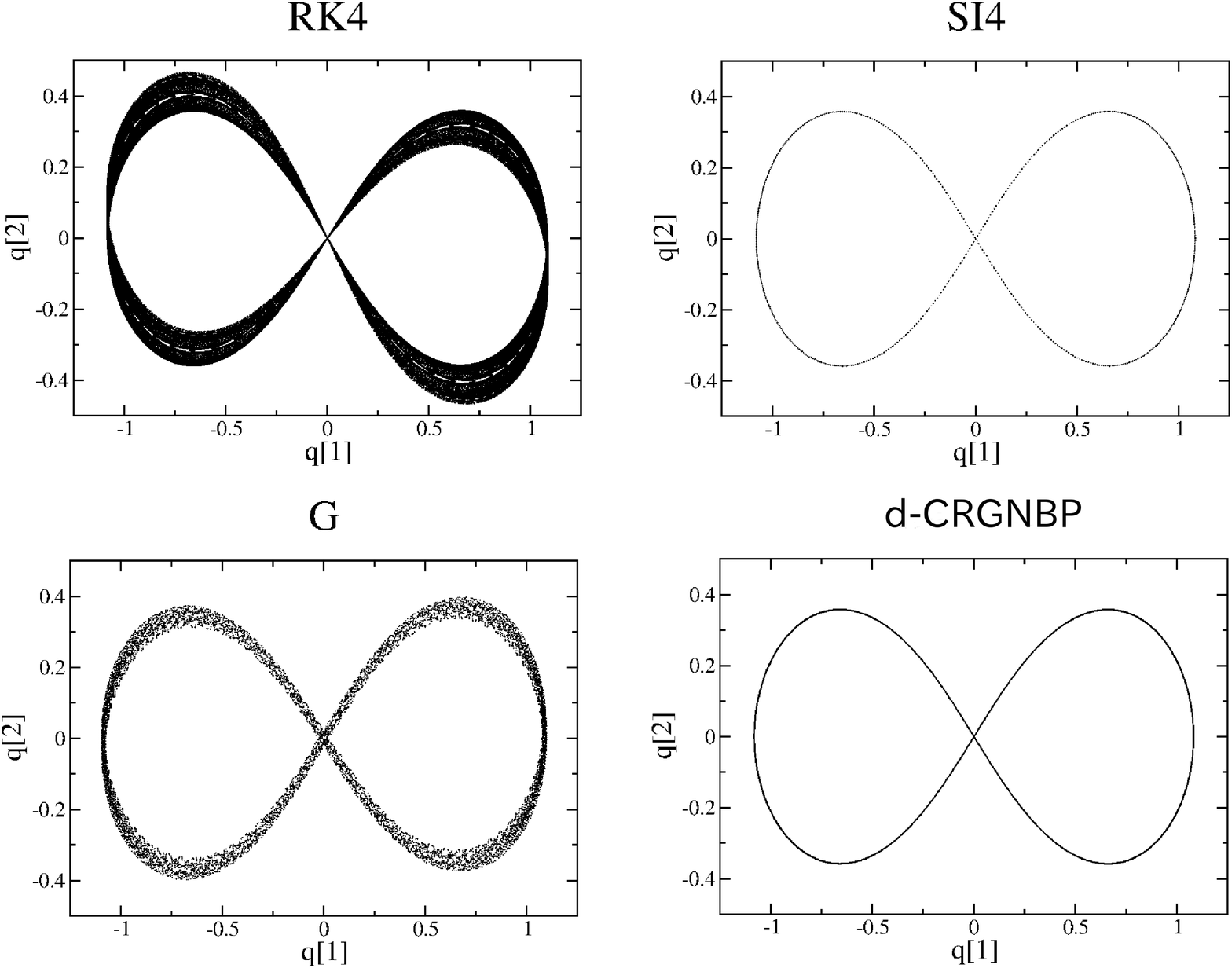}
\caption{
Orbits of particle $m_3$ for three-body figure-eight choreography
\citep{Chenciner, Simo-2000} computed by RK$4$, SI$4$, G, and
d-CRG$N$BP methods.
}
\end{figure}
\begin{figure}
\epsscale{.80}
\plotone{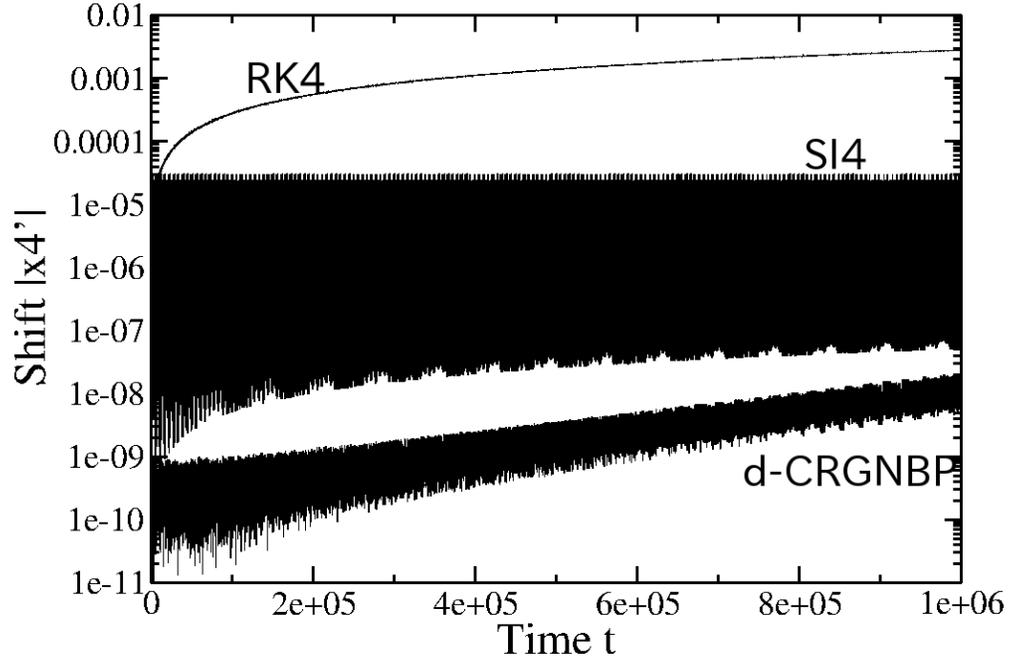}
\caption{
Particle $m_4$ shifts by $|\Delta \mathbf{x}'_4|$ from its initial position
$\mathbf{x}'_4 (0)$ in rotating frame $\mbox{O}-x'_{[1]} x'_{[2]}$
given by RK$4$, SI$4$, and d-CRG$N$BP methods.
The position $\mathbf{x}'_4(0)$ corresponds to an equilibrium point in the
G$4$BP.
The integrated position of $m_4$ corresponds to the initial condition
(\ref{equilibrium-CG4BP}).
}
\end{figure}
\begin{figure}
\epsscale{.80}
\plotone{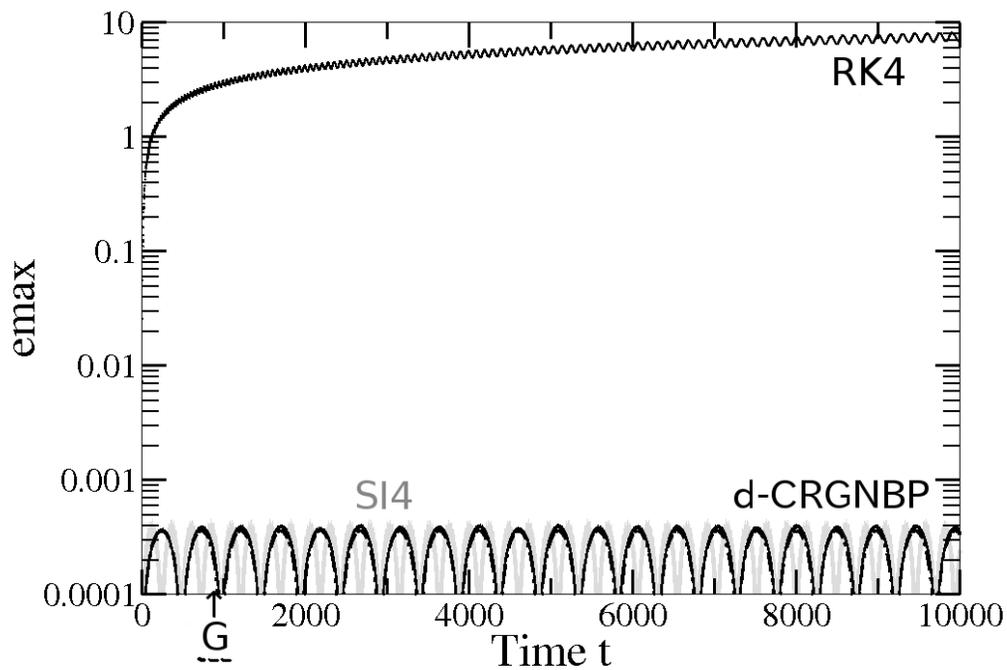}
\caption{
Maximum value of amplitudes of all shifts of particle $m_i$,
$|\mathbf{x}'_i (t) - \mathbf{x}'_i (0)|$, in the rotating frame
$\mbox{O}-x'_{[1]} x'_{[2]}$, $e_{\mbox{\footnotesize max}}$
given by RK$4$, SI$4$, G, and d-CRG$N$BP methods.
Each initial position $\mathbf{x}'_i (0)$ is an equilibrium point in the
$1+4$-body problem.
Each integrated value of $\mathbf{x}'_i (t)$ corresponds to the initial
 condition (\ref{Central-Configuration-1+4-BP}).
}
\end{figure}
\begin{figure}
\epsscale{1.0}
\plotone{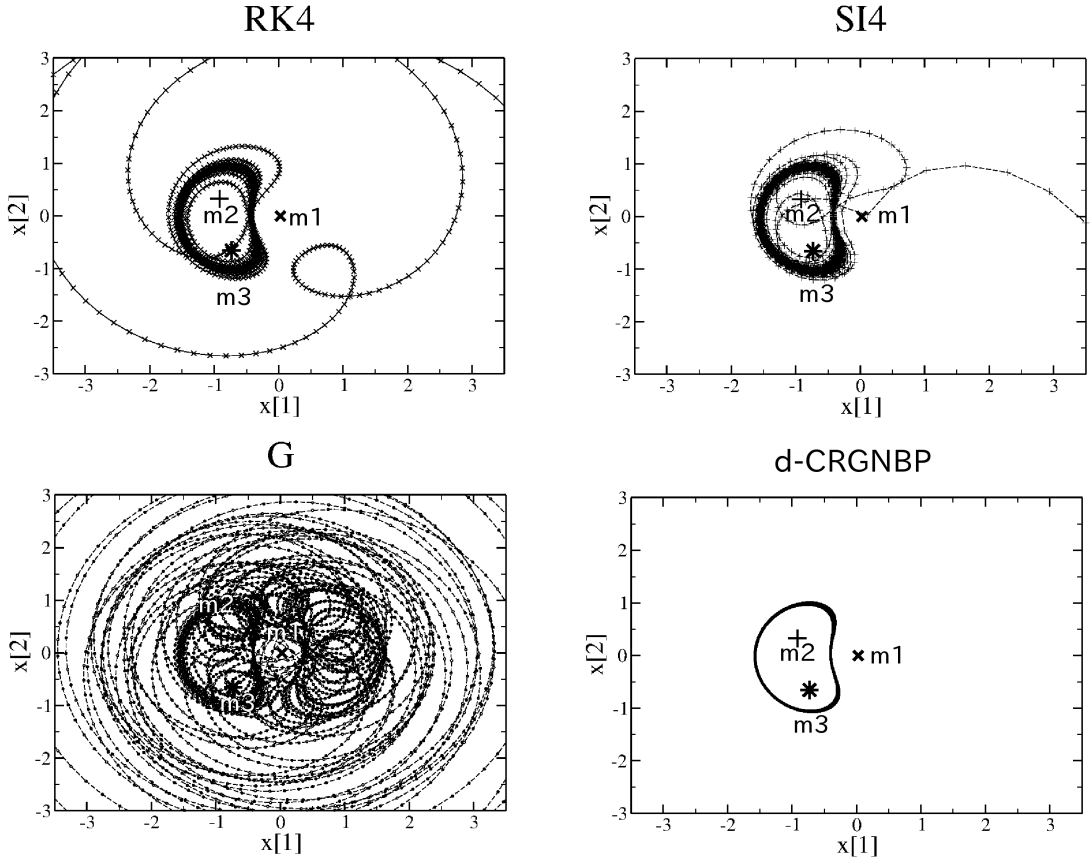}
\caption{
Orbits of primaries $m_1$, $m_2$, and $m_3$ in rotating frame
$\mbox{O}-x'_{[1]} x'_{[2]}$ given by RK$4$, SI$4$, G, and d-CRG$N$BP
methods.
The integrated orbit of particle $m_4$ corresponds to the initial condition
(\ref{Periodic-EG4BP}).
}
\end{figure}
\end{document}